\documentclass[12pt,a4paper]{article}
\usepackage{amsmath,amssymb,mathrsfs,framed,esint,slashed,mathabx,upgreek}
\usepackage[colorlinks]{hyperref}
\usepackage{color}
\usepackage[all]{xy}

\newtheorem{theorem}{Theorem}[section]

\newtheorem{remark}[theorem]{Remark}

\newcommand{\rem}[1]{}
\newcommand{\bLambda}{{\boldsymbol{\Lambda}}}

\newcommand{\de}{{\rm d}}

\newcommand{\bq}{{\boldsymbol{q}}}

\newcommand{\bp}{{\boldsymbol{p}}}

\newcommand{\bM}{{\mathbf{M}}}
\newcommand{\bX}{{\mathbf{X}}}

\newcommand{\bn}{{\mathbf{n}}}

\newcommand{\bs}{{\boldsymbol{s}}}
\newcommand{\bH}{{\boldsymbol{H}}}

\newcommand{\bz}{{\mathbf{z}}}

\newcommand{\bu}{{\boldsymbol{u}}}
\newcommand{\bnu}{{\boldsymbol{\nu}}}

\newcommand{\beq}{\begin{equation}}
\newcommand{\eeq}{\end{equation}}
\newcommand{\ben}{\begin{eqnarray}}
\newcommand{\een}{\end{eqnarray}}

\begin{document}

\title{Dynamics of mixed quantum-classical spin systems\footnote{Contribution to the Special Collection \href{https://iopscience.iop.org/journal/1751-8121/page/koopman-methods-in-classical-and-quantum-classical-mechanics}{``Koopman methods in classical and quantum-classical mechanics''} in the Journal of Physics A.}}
\author{Fran\c{c}ois Gay-Balmaz$^1$, Cesare Tronci$^{2,3}$ 
\\ 
\footnotesize
\it $^1$CNRS and \'Ecole Normale Sup\'erieure, Laboratoire de M\'et\'eorologie Dynamique, Paris, France
\\
\footnotesize
\it $^2$Department of Mathematics, University of Surrey, Guildford, United Kingdom
\\
\footnotesize
\it 
$^3$Department of Physics and Engineering Physics, Tulane University, New Orleans LA, United States}
\date{}

\maketitle

\begin{abstract}
Mixed quantum-classical spin systems have been proposed in spin chain theory and, more recently, in magnon spintronics. However, current models of quantum-classical dynamics beyond mean-field approximations typically suffer from long-standing consistency issues, and, in some cases, invalidate Heisenberg's uncertainty principle. Here, we present a fully Hamiltonian theory of quantum-classical spin dynamics that appears to be the first to ensure an entire series of consistency properties, including positivity of both the classical and the quantum density, so that Heisenberg's principle is satisfied at all times. We show how this theory may connect to recent energy-balance considerations in measurement theory and we present its Poisson bracket structure explicitly. After focusing on the simpler case of a  classical Bloch vector interacting with a quantum spin observable, we illustrate the extension of the model to systems with several spins, and restore the presence of orbital degrees of freedom.

\end{abstract}

{\footnotesize
\tableofcontents
}

\section{Introduction\label{sec:intro}}

The search for a mixed quantum-classical description of many-body quantum systems is motivated by the formidable challenges posed by the curse of dimensionality appearing in fully quantum approaches. Already at the level of Born-Oppenheimer theory \cite{BACaCaVe09,BoOp27,RaTr20}, it is common practice in molecular dynamics to approximate nuclei as classical particles while retaining a fully quantum electronic description \cite{FoHoTr19,HoRaTr21,Kapral,Tully}. Similar mixed quantum-classical approximations have also been proposed in quantum plasmas \cite{HuHeMa17}.

However, the interaction dynamics of quantum and classical degrees of freedom remains a challenging question since the currently  available models suffer from several consistency issues \cite{AgCi07}. In some cases \cite{Aleksandrov,Gerasimenko,boucher}, the Heisenberg principle is lost due to the fact that the quantum density matrix is allowed to change its sign. In some other cases \cite{Diosi,Hall}, the model does not reduce to uncoupled quantum and classical dynamics in the absence of a quantum-classical interaction potential. 
At a computational level, the most popular approach is probably  the Ehrenfest model, which is written as
\beq\label{Ehrenfest}
\partial_t D+\operatorname{div}\!\big(D\langle\bX_{\widehat{H}}\rangle\big)=0
\,,\quad\ 
i\hbar\big(\partial_t\psi+\langle\bX_{\widehat{H}}\rangle\cdot\nabla\psi\big)=\widehat{H}\psi
\,,\quad\,\text{with}\quad\,
\bX_{\widehat{H}}=\big(\partial_p\widehat{H},-\partial_q\widehat{H}\big).
\eeq
Here, $D(q,p)$ is the classical density, while $\psi(x;q,p)$ is a wavefunction depending on the quantum $x-$coordinate and parameterized by the classical coordinates $(q,p)$. Also,  $\widehat{H}(q,p)$ is a quantum Hamiltonian operator depending on $(q,p)$ and we have resorted to the usual notation $\langle \widehat{A}\rangle=\langle\psi|\widehat{A}(q,p)\psi\rangle$, where $\langle\psi_1|\psi_2\rangle=\int\psi^*_1(x)\psi_2(x)\,\de x$. In this setting, the matrix elements of the quantum density operator are given  as $\hat\rho(x,x')=\int D(q,p)\psi(x;q,p)\psi^*(x';q,p)\,\de q\de p$. Despite its wide popularity, the Ehrenfest model fails to reproduce realistic levels of decoherence, which is usually expressed in terms of the norm squared $\|\hat\rho\|^2$ of the density operator, a quantity also known as \emph{quantum purity}. This has led some authors \cite{AkLoPr14,CrBa18} to search for \emph{ad-hoc} corrections of the Ehrenfest model in order to retain more realistic decoherence levels.

In recent years, the authors proposed a theory of hybrid quantum-classical dynamics \cite{GBTr21,GBTr22} that captures correlations beyond the Ehrenfest model and still satisfies five of its important consistency properties \cite{boucher}: 1) the classical system is identified by a phase-space probability density at all times; 2) the quantum system is identified by a positive-semidefinite density operator $\hat\rho$ at all times; 3) the model is covariant under both quantum unitary transformations and classical canonical transformations; 4) in the absence of an interaction potential, the model reduces to uncoupled quantum and classical dynamics; 5) in the presence of an interaction potential, the {\it quantum purity} $\|\hat\rho\|^2$ is not a constant of motion (decoherence property). 

Blending Koopman wavefunctions in classical mechanics \cite{Koopman} with the geometry of prequantum theory \cite{Ko1970,Souriau,VanHove}, we formulated a quantum-classical model which was developed in two stages. First, we provided an early quantum-classical model \cite{BoGBTr19,GBTr20,MaRiTr23} that succeeded in satisfying only the properties 2)-5). 
Then, more recently, we upgraded this model in such a way that property 1) is also secured \cite{GBTr22,GBTr21}. This  upgrade results from applying a gauge principle on the original version of the model in order to ensure that classical phases are \emph{unobservable}, that is they do not contribute to measurable expectation values. Inspired by Sudarshan's work \cite{Marmo,Sudarshan}, this gauge principle naturally leads to crucial properties such as the characterization of entropy functionals and the Poincar\'e integral invariant, thereby extending the classical conserved quantity $\oint_{c(t)}p\de q$  to quantum-classical dynamics. Nevertheless, the model is nonlinear and its explicit form is rather intricate due to the appearance of a  gauge connection that extends Mead's non-Abelian  potential \cite{Me92} to phase-space. Upon defining the (Hermitian) operator-valued vector potential as $\boldsymbol{\widehat{\Gamma}}={i}[P,\nabla P]$, with $P(x,x';q,p)=\psi(x;q,p)\psi^*(x';q,p)$, one can write the model proposed in \cite{GBTr22,GBTr21} as 
\beq
\partial_t D+\operatorname{div}(D\boldsymbol{X})=0
\,,\qquad\ 
i\hbar(\partial_t\psi+\boldsymbol{X}\cdot\nabla\psi)=\widehat{\cal H}\psi,
\label{HybEq1}
\eeq
with
\beq
\boldsymbol{X}=
\langle\bX_{\widehat{H}}\rangle-\frac\hbar{2D}\operatorname{Tr}\!\Big( (DJ\boldsymbol{\widehat{\Gamma}})\cdot\nabla\bX_{{\widehat{H}}}
-\bX_{{\widehat{H}}}\cdot\nabla (DJ\boldsymbol{\widehat{\Gamma}})
\Big),
\qquad\qquad
J=\bigg(\begin{array}{cc}
0 & {1} \\
-{1} & 0
\end{array}
\bigg),
\label{HybEq2}
\eeq
and
\beq
\widehat{\mathcal{H}}= \widehat{H}+i\hbar\Big(\{P,{\widehat{H}}\}+\{\widehat{H},P\}-\frac{1}{2D}[\{ D,\widehat{H}\},P] \Big).
\label{HybEq3}
\eeq  
Thus, we conclude that the vector field $\boldsymbol{X}$ and the Hermitian generator $\widehat{\mathcal{H}}$ can be regarded as $\hbar-$modifications of the original Ehrenfest quantities. Notice that the minus sign before the trace in the expression of $\boldsymbol{X}$ fixes a typo in equation (6.10) of \cite{GBTr22}. While equations \eqref{HybEq1}-\eqref{HybEq3} appear hardly tractable at first sight, a direct calculation of $\operatorname{div}\!\boldsymbol{X}$ reveals that no gradients of order higher than two appear in the equations \eqref{HybEq1}. In addition,  the Hamiltonian/variational structure of this system reveals much of the features occurring in quantum-classical coupling. Thus, we consider the equations above as a platform for the formulation of simplified closure models that can be used in physically relevant cases. Trajectory-based numerical algorithms to implement equations  \eqref{HybEq1}-\eqref{HybEq3} are currently underway.

Despite this recent progress, current results only apply to classical  orbital degrees of freedom, that is canonical coordinates. However, several situations involve spin systems requiring a noncanonical treatment. Quantum-classical spin hybrid systems have been proposed recently in spintronics  \cite{RuKaUp22} as a way to control quantum impurity spins, although the general idea of quantum-classical spin hybrids \cite{Sergi2014} goes back to spin chain models in the theory of magnetism \cite{GeBoCoCuDr02}. In theoretical chemistry,  mixed quantum-classical spin dynamics may overcome the limitations of certain semiclassical models currently used to describe organic radical pairs \cite{Manolopoulos1}. In this context, electron spin-spin coupling generally requires a quantum treatment \cite{Manolopoulos2}, while at present the semiclassical description  applies to both nuclear and electronic spins.  Motivated by these investigations, here we provide a new formulation of  hybrid dynamics for correlated quantum-classical spin systems. Based on a variational formulation involving the phase-space of the rotation group, the new theory is obtained as a closure model that is made available by the underlying variational principle. The intricacies arising from the noncanonical spin structure   are dealt with by resorting to standard methods in reduction by symmetry \cite{HoScSt09,MaRa98}.

The paper proceeds as follows. After a brief summary of classical Bloch vector dynamics and the corresponding Liouville equation, Section \ref{sec:classicalspins} presents their underlying Hamiltonian and variational structures. Notice that this treatment is somewhat complicated by the fact that, as mentioned above, rotational motion  generally requires noncanonical coordinates. Then, we present two different Koopman formulations of classical Bloch dynamics: Koopman-von Neumann and Koopman-van Hove. These are eventually related by resorting to their underlying variational structures. Section \ref{sec:QChybs} is devoted to quantum-classical dynamics. After presenting a preliminary model based on the Koopman-van Hove construction, the discussion proceeds by applying a gauge principle that results in ensuring both a positive-definite classical density and a positive-semidefinite quantum density matrix. Then, the resulting model equations are  presented explicitly and shown to be highly nonlinear. Despite their formidable appearance, these equations do not involve gradients of order higher than two. In addition, Section \ref{sec:discussion} shows how various remarkable properties are inherited naturally from the underlying variational setting. After discussing some aspects concerning the energy balance, we consider the expectation values of hybrid quantum-classical observables. Also, we present the explicit Hamiltonian formulation of the model along with its Poisson bracket structure and the associated Casimir invariants. Finally, we discuss different augmentations and extensions to systems with multiple spins as well as orbital degrees of freedom.

\section{Classical spin evolution and variational structure\label{sec:classicalspins}}

\subsection{Classical Bloch vector dynamics and Liouville equation}

While spin is a typical quantum observable, the Bloch ball representation  offers an immediate classical analogue in terms of a vector evolving within the unit sphere. In the simplest case of a linear Hamiltonian function $H(\bn)=\bn\cdot\bH$, the Bloch vector dynamics is $\dot{\bn}=\bH\times\bn$ where  $\bH$ typically represents an external magnetic field.
Then, the spin vector is found as $\bs=\hbar\bn/2$, although here we will work with Bloch vectors to avoid the proliferation of unnecessary numerical factors. 

At the level of a single spin, in the absence of orbital degrees of freedom, the Bloch vector dynamics is essentially a classical precession. This observation may be used to formulate classical spin dynamics in the case of multiple spins $(\bn_1,\bn_2,\dots)$. In more generality, one may introduce a classical probability density $D(\bn_1,\bn_2,\dots)$ that is transported along the trajectories followed by the Bloch vectors. This produces a classical Liouville equation for spins.
For simplicity, here we will focus on a single classical spin described by the Bloch vector $\bn=2\bs/\hbar$. In this case, the Liouville equation for the density $D(\bn,t)$ reads
\beq\label{CLE}
\partial_tD=\nabla D\cdot\bn\times \nabla H
\,,
\eeq
where  $H(\bn)$ is now a generic Hamiltonian function.

The remainder of section will deal with the Hamiltonian and variational structures of  classical  spin systems and their formulation in terms of Koopman wavefunctions. In particular, the variational structure will  be exploited later on to devise a model for quantum-classical spin systems.

\subsection{Hamiltonian/variational structures of  classical  spin dynamics\label{sec:CLEq}}
We  begin this section by considering the dynamics of a single Bloch vector, which will provide the framework for the study of the classical Liouville density evolution.
The  precessional motion of the Bloch vector provides the most important example of a noncanonical Hamiltonian system with Lie-Poisson bracket 
\beq\label{LPB}
\{f,k\}=\bn\cdot\nabla f\times\nabla k
\,.
\eeq
Indeed, the relation $\dot\bn=\{\bn,H\}=\nabla H\times\bn$ recovers the classical vector dynamics previously introduced  for a linear Hamiltonian $H(\bn)=\bn\cdot\boldsymbol{H}$.
We notice that, while here we generally consider functions on the entire three-dimensional space, the restriction to functions on the unit sphere (that is, functions of unit vectors) does not change the  Poisson structure \eqref{LPB}; see Section \ref{sec:invariants} for further discussions on this point.

The corresponding variational structure is given by  Hamilton's variational principle on phase-space. If $R$ is a rotation matrix and $P$ is its conjugate momentum, one generally writes the variational principle as $\delta \int_{t_1}^{t_2} \big(  \operatorname{Tr}(P^T  \dot  R) - H(R, P) \big)\, \de t=0$ for arbitrary variations $ \delta R$, $ \delta P$. Here, $\operatorname{Tr}$ denotes the trace and the superscript $T$ denotes transpose. In the case of rotational symmetry, the Hamiltonian depends only on $\hat{n}=PR^T$ and one can use the usual isomorphism $\hat{v}_{lm}=-\epsilon_{lmn}v_n$ between three-dimensional vectors and skew-symmetric $3 \times 3$ matrices to write the entire action principle in terms of vectors. From this,  one may formulate a variational principle that is entirely expressed in terms of the Bloch vector and the angular frequency vector $\bnu$ associated to the matrix $\hat{\nu}=\dot R R^T$. 
Then, the equation $\dot\bn=\nabla H\times\bn$ may be obtained from the following variational principle:
\beq\label{EPSpin}
\delta\int_{t_1}^{t_2} \!\big(\bn\cdot\boldsymbol\nu-H(\bn)\big)\de t=0
\,,\qquad\text{ with }\qquad
H(\bn)=\bH\cdot\bn
\,.
\eeq
Here, $\delta\bn$ is arbitrary, while $\delta\boldsymbol\nu=\dot{\boldsymbol\xi}+\boldsymbol\xi\times\boldsymbol\nu$ with $\boldsymbol\xi$ an arbitrary time-dependent vector.  The latter expression for the variation $\delta\boldsymbol\nu$ of the angular frequency arises from the definition $\hat\nu=\dot{R}{R}^T$ upon denoting ${\hat\xi}=\delta{R}{R}^T$. In particular,  the variations $\delta\boldsymbol\nu$ give $\dot\bn=\boldsymbol\nu\times\bn$, while the variations $\delta\bn$ give $\boldsymbol\nu=\nabla H$.
This construction is valid for any Hamiltonian $H(\bn)$, while in the previous section we  considered only a linear expression for simplicity. Both the noncanonical Hamiltonian and variational structures arise from the theory of reduction by symmetry \cite{HoScSt09,MaRa98}, which however is beyond the scope of this article.

We now turn our attention to consider the classical Liouville equation \eqref{CLE}. The corresponding variational principle  has its roots in the Lagrangian trajectories from continuum theories and an early formulation is found in \cite{HoPuTr13}. If $\eta(R_0,P_0,t)$ is a Lagrangian trajectory on the phase-space corresponding to the rotation group, with components $\eta(R_0,P_0,t)=(\eta_R(R_0,P_0,t)  ,\eta_P(R_0,P_0,t))$, one starts with the variational principle $\delta \int_{t_1}^{t_2} \int \mathscr{D}_0\big(  \operatorname{Tr}(\eta_P^T  \dot{\eta}_R) - H(\eta_R  ,\eta_P) \big)\,\de^3 R_0 \de ^3 P_0 \de t=0$, for arbitrary variations $ \delta \eta_R$, $ \delta \eta_P$. Here $\mathscr{D}_0(R_0,P_0)$ is a reference density and $\de^3 R_0 \de ^3 P_0$ is the Liouville volume element on the phase space associated to the rotation group.
Upon changing coordinates $(R_0,P_0)\mapsto(R_0,\bn_0)$, the Lagrangian trajectory is written as $\tilde\eta(R_0,\bn_0)=(\tilde\eta_R(R_0,\bn_0)  ,\tilde{\boldsymbol\eta}(R_0,\bn_0))$ and similarly for ${\mathscr{D}}_0$, which becomes $\tilde{\mathscr{D}}_0(R_0,\bn_0)$. Then, the previous variational principle can be equivalently written as
\beq\label{intermvarprin}
\delta \int_{t_1}^{t_2} \int \tilde{\mathscr{D}}_0\big(  \operatorname{Tr}(\hat{\tilde\eta}^T \dot{\tilde\eta}_R\tilde\eta_R^{-1}) - H(\tilde\eta_R  ,\tilde{\boldsymbol\eta}) \big)\,\de^3 \mu_R\,\de^3 n_0 \,\de t=0,
\eeq
where $ \de^3\mu_R$ is the Haar volume element on the rotation group and $\de^3 n_0$ is the usual volume element on $ \mathbb{R} ^3$. Here, we used the fact that the Liouville volume form $\de^3 R_0 \de ^3 P_0$ on the phase space corresponds, up to a factor, to $\de^3 \mu_R\,\de^3 n_0 $, via the change of coordinates $(R_0,P_0)\mapsto(R_0,\bn_0)$; see \cite{Ha1997} for further details.

In the case of rotational symmetry, the Hamiltonian  depends only on $\tilde{\boldsymbol\eta}$. Then, the symmetry of the problem allows  restricting to a specific type of Lagrangian trajectories of the form $(\tilde\eta_R(R_0,\bn_0)  ,\tilde{\boldsymbol\eta}(R_0,\bn_0))=({\cal R}(\bn_0)R_0,\boldsymbol{\eta}(\bn_0))$. In geometric terms, this corresponds to selecting diffeomorphisms $\tilde\eta$ of phase space that are equivariant (or covariant, in more physical terms) with respect to rotations, that is $(\tilde\eta_R(R_0R',\bn_0)  ,\tilde{\boldsymbol\eta}(R_0R',\bn_0))= (\tilde\eta_R(R_0,\bn_0) R' ,\tilde{\boldsymbol\eta}(R_0,\bn_0))$ for any rotation matrix $R'$. 
Here, we will simply refer the reader to \cite{GBRa2008,GBViTr20} for further details on these geometric aspects.

Then, upon introducing $D_0=\int\tilde{\mathscr{D}}_0\,\de^3  \mu_R$, we define
\beq\label{LtEmap}
D(\bn,t)=\int \!D_0(\bn_0)\,\delta(\bn-{\boldsymbol\eta}(\bn_0,t))\,\de^3 n_0
\eeq
and $\hat\nu(\bn,t)=\dot{\cal R}(\bn_0,t){\cal R}^T(\bn_0,t)|_{\bn_0=\boldsymbol\eta^{-1}(\bn,t)}$, so that the  variational principle \eqref{intermvarprin} becomes
\beq\label{CLEVP}
\delta\int_{t_1}^{t_2} \ell(\boldsymbol{\cal X},\bnu,D)\de t=0
\,,\qquad\text{ with }\qquad
\ell(\boldsymbol{\cal X},\boldsymbol\nu,D)=
 \int D(\bn)  \big(\bn\cdot\boldsymbol\nu(\bn)-H(\bn)\big)\,\de^3 n
 \,.
\eeq
Here, the Lagrangian $\ell$ depends on the vector-valued scalar function $\bnu(\bn,t)$ (the angular frequency field), and formally also on the vector field $\boldsymbol{\cal X}(\bn,t)$ whose integral curves $\boldsymbol\eta$ satisfy $\dot{\boldsymbol\eta}(\bn,t)=\boldsymbol{\cal X}(\bn_0,t)|_{\bn_0=\boldsymbol\eta(\bn,t)}$. 
The definitions of $\boldsymbol{\cal X}$ and $D$ in terms of $\boldsymbol\eta$ produce the variations
\beq
\delta\boldsymbol{\cal X}=\partial_t\boldsymbol{\cal Y}+\boldsymbol{\cal X}\cdot\nabla\boldsymbol{\cal Y}-\boldsymbol{\cal Y}\cdot\nabla\boldsymbol{\cal X}
\,,\qquad\qquad
\delta D=-\operatorname{div}(D\boldsymbol{\cal Y})
\,,
\label{XDvar}
\eeq
where $\boldsymbol{\cal Y}$ is an arbitrary infinitesimal displacement satisfying $\delta{\boldsymbol\eta}(\bn,t)=\boldsymbol{\cal Y}(\bn_0,t)|_{\bn_0=\boldsymbol\eta(\bn,t)}$. In addition, the  frequency field $\bnu(\bn,t)$ identifies a generator of the $\bn-$dependent rotations governing the Lagrangian trajectories $\boldsymbol\eta$. From its definition in terms of $\boldsymbol\eta$ and ${\cal R}$, one finds
\beq
\delta\bnu=\partial_t\bLambda+\bLambda\times\bnu+\boldsymbol{\cal X}\cdot\nabla\bLambda-\boldsymbol{\cal Y}\cdot\nabla\bnu
\,,
\label{xivar}
\eeq
where $\hat\Lambda(\bn,t):=\delta{\cal R}(\bn_0,t){\cal R}^T(\bn_0,t)|_{\bn_0=\boldsymbol\eta^{-1}(\bn,t)}$ is an arbitrary quantity.

Since the particular function $\ell$ in \eqref{CLEVP} is actually independent of $\boldsymbol{\cal X}$, the $\boldsymbol{\cal Y}-$terms in the variational principle are produced only by the variations  $\delta D$ and $\delta\bnu$. These terms lead to the algebraic relation $
\bnu=\nabla H$. 
On the other hand, the $\bLambda-$terms give
$
\boldsymbol{\cal X}=\bnu\times\bn
$,
so that
\[
\boldsymbol{\cal X}=\nabla H\times\bn=:\bX_H
\,.
\]
Then, since the {\it Hamiltonian vector field $\bX_H$} is divergence-free, the equation $\partial_t D+\operatorname{div}(D\boldsymbol{\cal X})=0$ arising from the definition \eqref{LtEmap} returns \eqref{CLE}.

\subsection{Koopman wavefunctions}

This section presents the formulation of classical spin dynamics in terms of Koopman wavefunctions. As discussed in \cite{BoGBTr19,GBTr20,TrJo21}, the more popular Koopman-von Neumann (KvN) construction has a variant in prequantum theory that is known as Koopman-van Hove (KvH) formulation. The two essentially differ by a crucial phase factor that restores the information on classical phase dynamics in the KvH setting. This is a key ingredient in bridging across the quantum-classical divide, since phases are notably crucial in quantum evolution and thus the formulation of a hybrid model in which quantum and classical dynamics are regarded on an equal footing cannot  disregard the role of  phases in both sectors. Importantly, the extra phase factor affects the expression of the classical Liouville distribution in terms of the Koopman wavefunction. This expression is indeed different in the KvN and the KvH constructions, as explained in \cite{BoGBTr19,GBTr22}. Since these aspects have been discussed thoroughly elsewhere, here we will simply adapt them to the case of classical spin dynamics.

The KvN construction arises by simply writing the Liouville spin density in terms of a wavefunction $\chi( \mathbf{n} )$ as ${D}=|\chi|^2$. Then, the Liouville equation \eqref{CLE} leads to the KvN spin equation
\beq\label{KvNeq}
i\hbar\partial_t\chi=L_H\chi
\,,\qquad\text{ with }\qquad
L_H \chi =i\hbar\{H,\chi\}
\,.
\eeq
Here, the Hermitian operator  $L_H$  is known as {\it Liouvillian} and is expressed in terms of the Lie-Poisson bracket \eqref{LPB}. We notice that the KvN wavefunction is defined up to a phase factor, so that $\chi$ can be made real depending on convenience.

The KvH construction modifies the KvN equation by adding a phase term to the Liouvillian operator. The resulting modified Liouvillian in KvH theory is generally written as the Hermitian operator ${\cal L}_H=i\hbar\{H, \cdot   \}-\mathscr{L}$, where $\mathscr{L}$ is the Lagrangian corresponding to classical motion. In the case of classical spin dynamics, $\mathscr{L}$ is given as in  \eqref{EPSpin}, that is  $\mathscr{L}=\bn\cdot\nabla H-H$. Here, following the discussion in the previous section, we have replaced  $\boldsymbol\nu=\nabla H$ in  \eqref{EPSpin} so that the Lagrangian is entirely expressed in terms of the coordinate vector $\bn$. Then, the KvH equation for  classical spin dynamics reads
\beq
i\hbar\partial_t\chi=i\hbar\{H,\chi\}-(\bn\cdot\nabla H-H)\chi=:{\cal L}_H\chi
\,.
\label{KvHeq}
\eeq 
Notice that, in the particular case $H(\bn)=\bH\cdot\bn$, the phase term vanishes identically and the KvH equation returns KvN. However, here we will consider the  general case of an arbitrary Hamiltonian $H$. 
The variational structure of the spin KvH equation \eqref{KvHeq} is given by the usual Dirac-Frenkel principle
\beq
\delta\int^{t_2}_{t_1}\!L_{\rm KvH}(\chi,\partial_t\chi)\,\de t=0
\,,\qquad\quad
L_{\rm KvH}(\chi,\partial_t\chi)=\operatorname{Re}\!\int\!\chi^*(i\hbar\partial_t-{\cal L}_H)\,\chi\,\de^3 n
\,.
\label{VP-KvH}
\eeq
This readily identifies the expression of the total energy $\int\!\chi^*{\cal L}_H\chi\,\de^3 n=\int\! H D\,\de^3 n$, so that the classical spin distribution $D(\bn)$ reads
\begin{align}
D=&\, |\chi|^2+\operatorname{div}(\bn|\chi|^2)+\hbar\operatorname{Im}\{\chi^*,\chi\}
\label{KvHmomap}
\,.
\end{align}
Then, one verifies that \eqref{KvHmomap} indeed satisfies the classical Liouville equation \eqref{CLE}.
We notice that the KvH construction differs substantially from the standard KvN theory and this difference arises from the phase factor, which in turn reflects in a different expression of the density $D$ in \eqref{KvHmomap}, compared to the density $D=| \chi | ^2 $ associated to \eqref{KvNeq}.

One may ask whether the KvN and KvH constructions are somehow related. In canonical coordinates $(q,p)$, this relation was recently unfolded in terms of the variational structure underlying the KvH equation \cite{GBTr22}. However, in noncanonical coordinates this relation is more subtle and requires resorting to the full phase-space variables $(R,\bn)$. This is the topic of the next section.

\subsection{The relation between KvH and KvN for  spin systems\label{sec:KvH-KvN}}

In this section we will address the question whether KvN dynamics can be somehow obtained from the more complete KvH theory. As anticipated, this requires a full phase-space approach in terms of both the classical vector $\bn$ and the rotational matrix $R$. In terms of canonical coordinates $(R,P)$ on phase-space associated to the rotation group, one has $\hat{n}=PR^T$. In terms of $(R,\bn)$, the equations of motion of a classical trajectory read
\beq
(\dot{R},\dot\bn)=\widetilde{X}_H({R},\bn)
\,,\qquad\text{ with }\qquad
\widetilde{X}_H({R},\bn):=(\widehat{\nabla H}R,\nabla H\times\bn)
\,.
\label{EQPScompl}
\eeq
Notice that in the system  \eqref{EQPScompl} the equation for $R$ is enslaved to that for $\bn$, so that $R$ is simply the instantaneous rotational state identified by the vector $\bn$ and no new information is actually added. This is due to the fact that the Hamiltonian $H$ depends only on $\bn$.

Then, the KvH equation obtained by extending \eqref{KvHeq} to phase space wavefunctions $\tilde\chi(R, \mathbf{n} )$ reads
\beq
i\hbar\partial_t\tilde\chi=-i\hbar \widetilde{X}_H \cdot\nabla\tilde\chi-(\bn\cdot\nabla H-H)\tilde\chi
\,,
\label{tildeKvH}
\eeq
which indeed reduces to \eqref{KvHeq} upon setting $\partial_R\tilde\chi=0$. As in the previous case, the right hand side now defines an Hermitian operator on phase space wavefunctions.  
To avoid proliferation of notation, here we have extended the dot product  as follows: $A\cdot B:=\operatorname{Tr}(A^T_RB_R)+\boldsymbol{A}\cdot\boldsymbol{B}$ for any $A=(A_R,\boldsymbol{A})$ and $B=(B_R,\boldsymbol{B})$. In addition, we have denoted $\nabla=(\partial_R,\partial_\bn)$ so that $\nabla H=(0,\partial_\bn H)$. We will perform a slight abuse of notation by confusing $\nabla H$ with $\partial_\bn H$.

Before proceeding, we emphasize that the Hermitian property of the right hand side of \eqref{tildeKvH} follows by recalling that the inner product on phase-space wavefunctions is computed with the volume form ${\rm d}^3 \mu_R\, {\rm d}^3n$ in the $(R, \mathbf{n} )$ description, which corresponds, up to a factor, to the Liouville form ${\rm d}^3 R \, {\rm d}^3P$ in the canonical $(R,P)$ description \cite{Ha1997}. Indeed, this ensures the usual permutation properties of Poisson brackets under the integral symbol, which then lead to the hermiticity of the right hand side of \eqref{tildeKvH}.

In the present setting, the  variational principle analogue to \eqref{VP-KvH} reads
\begin{equation}\label{VP_KvH_extended} 
\delta \int_{t_0}^{t_1} \operatorname{Re}\!\int\!\chi^* \big(i\hbar\partial_t\tilde\chi + i\hbar \widetilde{X}_H \cdot\nabla\tilde\chi + ({\cal A}\cdot\widetilde{X}_H - H)\tilde\chi\big)\de^3 \mu_R\, \de^3 n\,\de t.
\end{equation} 
For later convenience, here we have defined ${\cal A}=(\hat{n}R,0)$, so that 
\[
\bn\cdot\nabla H=\operatorname{Tr}(\hat{n}^T\widehat{\nabla H})={\cal A}\cdot\widetilde{X}_H
\,.
\] 
In particular, the differential form $\hat{n}R\cdot\de R=P\cdot\de R$ corresponds to the usual elementary action $p\de q$ in canonical coordinates.
Upon using the polar form $\tilde\chi=\sqrt{\tilde\varrho}\,e^{i\tilde{S}/\hbar}$, equation  \eqref{tildeKvH} leads to
\beq
\partial_t\tilde\varrho+\widetilde{X}_H \cdot\nabla\tilde\varrho=0
\,,\qquad\qquad
\partial_t\tilde{S}+\widetilde{X}_H \cdot\nabla \tilde{S}=\bn\cdot\nabla H-H
\,,
\label{MadEq1}
\eeq
and the variational principle \eqref{VP_KvH_extended} becomes
\beq
\delta\int^{t_2}_{t_1}\!\int\!\tilde{\varrho}\big(\partial_t \tilde{S}+ \widetilde{X}_H \cdot(\nabla \tilde{S}-{
\cal A})+H\big)\,\de^3 \mu_R\,\de^3 n\,\de t=0
\,.
\label{VP-KvH2}
\eeq

At this point, we consider the first equation in \eqref{MadEq1} and let $\tilde\varrho$ be transported along the integral curves of a vector field $\widetilde{\cal X}(R,\bn,t)$.   In other words, we consider the natural extension of \eqref{LtEmap} and let $\tilde\varrho$ evolve as
\beq
\tilde\varrho(R,\bn,t)=\int \!\tilde\varrho_0(R_0,\bn_0)\,\delta((R,\bn)-{\tilde\eta}(R_0,\bn_0,t))\,{\de^3 \mu_R\,}\de^3 n_0
, 
\label{tildeLtE}
\eeq
for some Lagrangian path $\tilde{\eta}$ such that $\partial_t{\tilde{\eta}}=\widetilde{\cal X}(\tilde{\eta},t)$. Then,  making use of the resulting relation $\partial_t\tilde\varrho+\operatorname{div}(\tilde\varrho\widetilde{\cal X})=0$ after integrating by parts the first term in \eqref{VP-KvH2}, we obtain the reduced variational principle
\beq
\delta\int^{t_2}_{t_1}\!\int\!\tilde\varrho\big(\widetilde{\cal X}\cdot\nabla \tilde{S}- \widetilde{X}_H \cdot(\nabla \tilde{S}-{\cal A}) - H\big)\,\de^3 \mu_R\,\de^3 n\,\de t=0,
\,.
\label{VP-KvH3}
\eeq
At this point, since the variations of $\tilde{\eta}$ are arbitrary, one gets the variations  $\delta\widetilde{\cal X}=\partial_t\widetilde{\cal Y}+\widetilde{\cal X}\cdot\nabla\widetilde{\cal Y}-{\cal Y}\cdot\nabla\widetilde{\cal X}$ and $\delta\tilde\varrho=-\operatorname{div}(\tilde\varrho\widetilde{\cal Y})$ and \eqref{VP-KvH3} yields the equations \eqref{MadEq1}, with the phase $\tilde{S}$ defined up to an irrelevant time-dependent number.

Without loss of generality, we follow the process from Section \ref{sec:CLEq} and consider Lagrangian paths $\tilde{\eta}(R,\bn)$ of the form $\tilde{\eta}(R,\bn)=({\cal R}(\bn)R,\boldsymbol{\eta}(\bn))$. This leads to writing the  vector field $\widetilde{\cal X}$ as
\beq
\widetilde{\cal X}(R,\bn)=(\hat\nu(\bn)R,\boldsymbol{\cal X}(\bn))
,
\label{equivVF}
\eeq 
where $\hat\nu(\bn,t)=\dot{\cal R}(\bn_0,t){\cal R}^t(\bn_0,t)|_{\bn_0=\boldsymbol\eta^{-1}(\bn,t)}$ is a $\bn-$dependent skew-symmetric matrix as in  Section \ref{sec:CLEq}. This  type of vector field reflects the expression of $\widetilde{X}_H$ in \eqref{EQPScompl} and takes \eqref{VP-KvH3} into
\beq
\delta\int^{t_2}_{t_1}\!\int\!\big(\boldsymbol{\cal X}\cdot\bM+\bnu\cdot\boldsymbol{\cal M}+  (\varrho\bn+\bM\times\bn-\boldsymbol{\cal M})\cdot\nabla H-\varrho H\big)\,\de^3 n\,\de t=0
\,,
\label{VP-KvH4}
\eeq
where the expressions of $\delta\boldsymbol{\cal X}$ and $\delta\bnu$ are found exactly as in \eqref{XDvar} and \eqref{xivar}. Notice that the integration is now over $ {\rm d} ^3n$ only. In \eqref{VP-KvH4} we have defined $\varrho( \mathbf{n} )$, $\bM( \mathbf{n} )$, and $\boldsymbol{\cal M}( \mathbf{n} )$ as
\beq
\varrho:=\int\!\tilde\varrho\,\de^3 \mu_R
\,,\qquad\ 
\bM:=\int\!\tilde\varrho\,\partial_\bn\tilde{S}\,\de^3 \mu_R
\,,\qquad\ 
\widehat{\cal M}:=\int\!\tilde\varrho\,\partial_R\tilde{S}R^T\de^3 \mu_R
\,.
\label{definitions}
\eeq

Due to the special form of $\tilde\eta$,  we observe from \eqref{tildeLtE} that $\varrho$ evolves exactly the same as $D$ in  \eqref{LtEmap} so that $\delta \varrho=-\operatorname{div}(\varrho\boldsymbol{\cal Y})$, while $\delta \bM$ and $\delta \boldsymbol{\cal M}$ are arbitrary since  so is $ \delta \tilde S$. From the variations $\delta \bM$ and $\delta \boldsymbol{\cal M}$ in \eqref{VP-KvH4} one gets $\boldsymbol{\cal X}= \nabla H \times \mathbf{n} $ and $\bnu= \nabla H$. Then, collecting  the terms proportional to $\boldsymbol{\Lambda }$ and $\boldsymbol{\cal Y}$ arising by taking the variations $\delta\boldsymbol{\cal X}$, $\delta\bnu$, and $ \delta \varrho$,  one gets
\beq
\partial_t\boldsymbol{\cal M}+\bX_H\cdot\nabla\boldsymbol{\cal M}=\nabla H\times\boldsymbol{\cal M}
\,,\qquad\quad
\partial_t\bM+\bX_H\cdot\nabla\bM+\nabla\bX_H\cdot\bM=\varrho\nabla(\bn\cdot\nabla H-H)-\nabla\bnu\cdot \boldsymbol{\cal M}
\,,
\label{M-eqns}
\eeq
as well as $\partial_t\varrho+ \bX_H\cdot\nabla \varrho=0$  with $\bX_H:=\nabla H\times\bn$. 
While these equations follow directly from the KvH equation \eqref{tildeKvH} for $\tilde\chi=\sqrt{\tilde\varrho}\,e^{i\tilde{S}/\hbar}$ by using  \eqref{definitions}, they have been obtained here in a more systematic way by using the variational principle \eqref{VP-KvH4}.
At this point, we notice that the first equation above has two solutions, each leading to a different Koopman construction.

\begin{enumerate}
\item {\bf Exact solution $\boldsymbol{\cal M}=0$ (KvH).} This is the case, for example, of a wavefunction $\tilde{\chi}$ in \eqref{tildeKvH} of the type $\tilde\chi(R,\bn)=\chi(\bn)A(R)$, where $A$ is a real amplitude while $\chi(\bn)$ is a complex-valued wavefunction. In this case, we obtain 
\begin{equation}\label{M_0} 
\partial_t\varrho+\bX_H\cdot\nabla \varrho=0
\,,\qquad\qquad
\partial_t\bM+\bX_H\cdot\nabla\bM+\nabla\bX_H\cdot\bM=\varrho\nabla(\bn\cdot\nabla H-H)
\,.
\end{equation} 
Then, writing  $\bM=\varrho\nabla S$ and $\varrho=|\chi|^2$, for some Koopman wavefunction $\chi(\bn)=\sqrt{\varrho(\bn)}e^{iS(\bn)/\hbar}$, returns the Madelung dynamics corresponding to the original KvH equation \eqref{KvHeq}. 
Indeed, \eqref{M_0} may be obtained directly from \eqref{KvHeq} upon using the polar form of $\chi$.

\item {\bf Exact solution $\boldsymbol{\cal M}=\varrho\bn$ (KvN).} As it may be verified directly, one has $\partial_t(\boldsymbol{\cal M}/\rho)+\bX_H\cdot\nabla(\boldsymbol{\cal M}/\rho)=\nabla H\times(\boldsymbol{\cal M}/\rho)$, so that the relation $\boldsymbol{\cal M}=\varrho\bn$ identifies an exact solution of the first equation in \eqref{M-eqns}. Moreover, in this case, we obtain $\bnu=\nabla H$ and the resulting $\bM-$equation  decouples entirely. As a result, we have 
\[
\partial_t\varrho+\bX_H\cdot\nabla \varrho=0
\,,\qquad\qquad
\partial_t\bM+\bX_H\cdot\nabla\bM+\nabla\bX_H\cdot\bM=0
\,.
\]
Notice that, while this solution is allowed by the equations \eqref{M-eqns}, it does not generally correspond to a solution of the original equation \eqref{KvNeq}. Nevertheless,   writing  $\bM=\varrho\nabla {\cal S}$ and $\varrho=|\chi|^2$, for some $\chi$, returns the Madelung dynamics corresponding to the original  KvN equation \eqref{KvNeq}. Also, setting $\boldsymbol{\cal M}=\varrho\bn$ and $\bM=0$ takes the variational principle \eqref{VP-KvH4} into the form \eqref{CLEVP} (upon replacing $D\to\varrho$), thereby returning the classical Liouville equation. Notice that, in this case, the phase is completely eliminated from the dynamics since the variational principle \eqref{CLEVP} is entirely phase-independent.

\end{enumerate}

\begin{remark}[From KvH to KvN]As pointed out above, the relation $\boldsymbol{\cal M}=\varrho\bn$ cannot be realized as arising from an exact solution of the general KvH equation \eqref{tildeKvH}. Nevertheless, upon writing  $\bM=\varrho\nabla {\cal S}$, this relation does lead us to the KvN formulation based on a wavefunction $\chi=\sqrt{\varrho}\,e^{i{\cal S}/\hbar}$, where $\varrho$ satisfies the first in \eqref{M_0} and analogously for $\cal S$.  A similar situation also occurs in canonical coordinates $(q,p)$. We showed in \cite{GBTr22} how a more direct relation between KvN and KvH dynamics may be identified by resorting to mixtures of Koopman wavefunctions. This point, however, is beyond our present scope.
\label{remark}
\end{remark}

In summary, we have shown that both the KvH equation \eqref{KvHeq} and the KvN equation  \eqref{KvNeq} may be found by suitable specializations of the variational principle \eqref{VP-KvH4} originally underlying the general KvH equation \eqref{tildeKvH} on the full phase-space with coordinates $(R,\bn)$.
In the remainder of this paper, we will exploit this setting to devise models for the coupled dynamics of hybrid quantum-classical spin systems.

\section{Quantum-classical spin hybrids\label{sec:QChybs}}

This part of the paper is entirely devoted to formulate a theory of quantum-classical spin dynamics. Strongly based on our previous considerations in the purely classical setting, a new model will be presented in two stages. First, we will present a variant that is constructed directly from the KvH formalism. This variant seems to satisfy a series of relevant properties except for the fact that the expression of the classical density is generally unsigned and it has not yet been proved to remain positive in time. In order to address this point, we will move on to the second stage in which the classical density will be made positive definite by applying a gauge principle that makes classical phases unobservable. For the case of classical orbital degrees of freedom, the resulting model was presented in \cite{GBTr22,GBTr21}, while here we focus on spin systems and deal with the intricacies arising from noncanonical classical coordinates.

\subsection{Quantum-classical KvH construction\label{sec:hybmod1}}

A first dynamical model of quantum-classical spin systems may be obtained by simply starting with the KvH equation \eqref{KvHeq} for two classical spins, $\bn$ and $\bn'$, and then quantize one of them. In general, this quantization process  follows from sophisticated techniques such as geometric quantization, which is entirely based on the prequantum KvH construction. This treatment is highly technical and is left  for other venues. In this paper, we will simply state the final result, which is obtained by first making the replacement $H\to{\sf H}$ in the two-particle KvH equation, where ${\sf H}(\bn)$ is a function of the classical Bloch vector with values in the space of Hermitian $2\times 2$ matrices. Notice that we have used sans serif fonts instead of the usual hat notation  to minimize possible confusion with the hat map identifying $3\times 3$ skew-symmetric matrices. In addition, we  also replace $\chi(\bn,\bn')\to\Upsilon(\bn)$, where $\Upsilon$ is a classical wavefunction with values in the space $\Bbb{C}^2$ of two-dimensional complex vectors. Then, one is left with the following quantum-classical KvH equation for the Pauli spinor $\Upsilon(\bn,t)$:
\beq
i\hbar\partial_t\Upsilon=i\hbar\{{\sf H},\Upsilon\}-(\bn\cdot\nabla {\sf H}-{\sf H})\Upsilon=:{\cal L}_{{\sf H}}\Upsilon
\,,
\label{HybKvHeq}
\eeq
where we recall the Lie-Poisson bracket notation \eqref{LPB}, so that $\{{\sf H},\Upsilon\}_a=\bn\cdot\nabla{\sf H}_{ab}\times\nabla\Upsilon_b$.  Explicitly, since we are dealing with only one quantum spin, we may write ${\sf H}(\bn) ={H}(\bn)+\bH(\bn)\cdot\widehat{\boldsymbol\sigma}$, where we have used the Pauli matrix notation $\widehat{\boldsymbol\sigma}=(\widehat{\sigma}_x,\widehat{\sigma}_y,\widehat{\sigma}_z)$ carrying the hat symbol. Notice that in the present paper we restrict to consider only two-level quantum systems and the extension to more general spin numbers is straightforward.

Equation \eqref{HybKvHeq} is the immediate extension of the quantum-classical wave equation proposed in \cite{BoGBTr19} for the case of canonical hybrid systems. In particular, since the hybrid Liouvillian ${\cal L}_{{\sf H}}$ is Hermitian, equation \eqref{HybKvHeq} possesses a variational formulation analogous to \eqref{VP-KvH}. In particular, the quantum-classical variational principle reads
\beq
\delta\int^{t_2}_{t_1}\!\operatorname{Re}\!\int\langle\Upsilon|(i\hbar\partial_t-{\cal L}_{{\sf H}})\Upsilon\rangle\,\de^3 n\,\de t=0
\,.
\label{VP-HybKvH}
\eeq
Here, the bracket $\langle\cdot  |\cdot  \rangle$ denotes the inner product in $\Bbb{C}^2$, so that $\langle \Upsilon_1 |\Upsilon_2 \rangle=\Upsilon_1^\dagger \Upsilon_2$ and $\|\Upsilon\|^2=\langle \Upsilon |\Upsilon \rangle$. Analogously to the classical case, one may integrate by parts to write the total energy as $\int\langle\Upsilon|{\cal L}_{{\sf H}}\Upsilon\rangle\,\de^3 n=\operatorname{Tr}\int{{\sf H}}{\cal D}\,\de^3 n$, where
\beq\label{D_hat}
{\cal D}=\Upsilon\Upsilon^\dagger+\operatorname{div}(\bn\Upsilon\Upsilon^\dagger)+i\hbar\{\Upsilon,\Upsilon^\dagger\}
\eeq
defines a hybrid von-Neumann operator mimicking the usual density matrix. In particular, ${\cal D}$ is a distribution-valued Hermitian operator acting  on  two-dimensional complex vectors.
Then, the quantum density matrix of the quantum subsystem is obtained as
\beq\label{rho_q}
\hat\rho_q=\int {\cal D}\,\de^3n=\int \Upsilon\Upsilon^\dagger\,\de^3n
\,,
\qquad\ \text{which satisfies}\qquad\ 
i\hbar\frac{\de \hat\rho_q}{\de t}=\int[{\sf H},{\cal D}]\,\de^3n
\,.
\eeq
We notice that, in analogy with the  results in \cite{BoGBTr19}, this quantity is again positive-definite thereby retaining the Heisenberg uncertainty principle in the quantum sector. In addition, the classical distribution is found as
\begin{equation}\label{rho_c} 
\rho_c=\operatorname{Tr} {\cal D}=
\|\Upsilon\|^2+\operatorname{div}(\bn\|\Upsilon\|^2)+\hbar\operatorname{Im}\{\Upsilon^\dagger,\Upsilon\}
\,,
\qquad\ \text{which satisfies}\qquad\ 
{\partial_t \rho_c}=\operatorname{Tr}\{{\sf H},{\cal D}\}
\,.
\end{equation} 
This expression is not positive-definite in general, although, by using methods analogous to those in \cite{GBTr20}, one can prove that hybrid Hamiltonians  depending only on one of the three Pauli matrices lead to a sign-preserving evolution for $\rho_c$. 

Being the natural extension to spin systems of the quantum-classical wave equation in \cite{BoGBTr19}, equation \eqref{HybKvHeq} inherits all the consistency properties already appearing in the case of canonical coordinates. If the positivity of the classical distribution stands as an additional requirement, however, the present model may need to be modified in such a way that this condition is satisfied at all times by construction. In the case of canonical coordinates, this upgrade of the quantum-classical wave equation was formulated in \cite{GBTr22,GBTr21} by blending variational methods with wavefunction factorization techniques currently used in chemical physics \cite{AbediEtAl2012}. This approach will be extended to quantum-classical spin systems in the next section, where we will show how ensuring both quantum and classical positivity leads to a nonlinear hybrid model.

\subsection{Nonlinear model I:  factorization and variational principle}

Upon following the discussion in Section \ref{sec:KvH-KvN}, our point of departure in the formulation of a positivity-preserving model consists in a slight generalization of the hybrid KvH equation \eqref{HybKvHeq}. In particular, we introduce a hybrid wavefunction $\tilde\Upsilon(R,\bn)$ on the full phase-space  and satisfying the  equation
\beq
\label{GenHybKvH}
i\hbar\partial_t\tilde\Upsilon=-i\hbar \widetilde{X}_{\sf H} \cdot\nabla\tilde\Upsilon-(\bn\cdot\nabla {\sf H}-{\sf H})\tilde\Upsilon
\,,\qquad\text{ where }\qquad
\widetilde{X}_{{\sf H}}({R},\bn):=(\widehat{\nabla {\sf H}}R,\nabla {\sf H}\times\bn)
\eeq
is an operator-valued vector field on the full phase-space.
In order to follow the steps in Section \ref{sec:KvH-KvN}, we  need to find a way to extract information on the classical phase. Indeed, the latter plays a crucial role in determining the particular Koopman construction that will be used. However, at this stage it appears difficult to extract  the classical phase from the hybrid wavefunction $\tilde\Upsilon(R,\bn)$, since the latter seems to blend together quantum and classical properties. The solution to this problem lies in a convenient factorization of the hybrid wavefunction, which will be the focus of the present discussion.

We use the factorization ansatz \cite{AbediEtAl2012}
\beq
\tilde\Upsilon(R,\bn)=\tilde\chi(R,\bn)\psi(\bn)
\,,\qquad\text{ with }\qquad
\|\psi(\bn)\|^2=1
\,.
\label{ExFact}
\eeq
Here, the factor $\psi(\bn)$ is a scalar function with values in $\Bbb{C}^2$, while $\tilde\chi(R,\bn)$ is simply a complex-valued wavefunction. The extra normalization  condition in \eqref{ExFact}  says that $\psi$ represents a  normalized quantum spin state that is parameterized by the classical vector $\bn$. In the  case when $\psi$ depends also on $R$ the factorization ansatz \eqref{ExFact} identifies an exact solution of \eqref{GenHybKvH} as long as $\tilde\chi$ vanishes nowhere. Instead of pursuing this more general case, here we simplify the treatment by restricting to consider a $\psi$ depending only on $\bn$. This allows us to retain quantum-classical correlations while treating the rotational coordinate only within the classical sector. As we will see, the presence of the classical wavefunction $\tilde\chi(R,\bn)$, which now encodes the classical density and phase, allows for convenient manipulations in the same spirit as  in Section \ref{sec:KvH-KvN}.

Then, replacing \eqref{ExFact} in the variational principle  
\[
\delta\int^{t_2}_{t_1}\!\operatorname{Re}\!\int\!\big\langle\tilde\Upsilon\big|\big(i\hbar\partial_t+i\hbar \widetilde{X}_{\sf H} \cdot\nabla+(\bn\cdot\nabla {\sf H}-{\sf H})\big)\tilde\Upsilon\big\rangle\,\de^3\mu_R\,\de^3 n\,\de t=0
\]
underlying the general equation \eqref{GenHybKvH}, writing $\tilde\chi=\sqrt{\tilde\varrho}\,e^{i\tilde{S}/\hbar}$, and dealing with the term $\int\tilde{\varrho}\partial_t \tilde{S}$\linebreak${ \de^3\mu_R}\,\de^3 n$ as in Section \ref{sec:KvH-KvN} yields
\beq
\delta\int^{t_2}_{t_1}\!\int\!\tilde\varrho\big(\nabla \tilde{S}\cdot(\widetilde{\cal X}-\langle\widetilde{X}_{\sf H}\rangle) +\langle\psi,(i\hbar \partial _t +i\hbar \widetilde{X}_{\sf H} \cdot\nabla)\psi\rangle+\langle{\cal A}\cdot\widetilde{X}_{\sf H}-{\sf H}\rangle\big)\,\de^3\mu_R\,\de^3 n\,\de t=0
\,,
\label{VP-HybKvH2}
\eeq
where we have introduced the notation
$
\langle\cdot ,\cdot \rangle=\operatorname{Re}\,\langle\cdot |\cdot\rangle
$ and $
\langle{\sf A}\rangle = \langle\psi,{\sf A}\psi\rangle
$.
Notice that the vector field $\widetilde{\cal X}$ in \eqref{VP-HybKvH2} is defined in exactly the same way as the vector field appearing in the variational principle \eqref{VP-KvH3}. Also, we  recall the notation ${\cal A}=(\hat{n}R,0)$ so that $\bn\cdot\nabla {\sf H}={\cal A}\cdot\widetilde{X}_{\sf H}=\operatorname{Tr}(\hat{n}^{T}\widehat{\nabla \sf H})$, where $\operatorname{Tr}$ denotes the partial trace involving only the classical degrees of freedom. Moreover, we recall the variations $\delta\widetilde{\cal X}=\partial_t\widetilde{\cal Y}+\widetilde{\cal X}\cdot\nabla\widetilde{\cal Y}-\widetilde{\cal Y}\cdot\nabla\widetilde{\cal X}$ and $\delta\tilde\varrho=-\operatorname{div}(\tilde\varrho\widetilde{\cal Y})$ to be used in \eqref{VP-HybKvH2}, while $\delta \tilde{S}$ and $\delta\psi$ are arbitrary.
Notice that at this stage, from the expression $\tilde\varrho\,\langle\psi,i\hbar \widetilde{X}_{\sf H} \cdot\nabla\psi\rangle+\tilde\varrho\,\langle({\cal A}-\nabla \tilde{S})\cdot\widetilde{X}_{\sf H}-{\sf H}\rangle$ of the total energy density, we conclude that the classical density   $\rho_c(\bn)$ acquires the form
\beq\label{classdens2}
\rho_c=\varrho+\operatorname{div}\!\big(\varrho\bn-\boldsymbol{\cal M}-\bn\times(\bM+\varrho\boldsymbol{\cal A}_B)\big),
\eeq
where we recall the definitions \eqref{definitions} and we have defined the Berry connection
\beq
\boldsymbol{\cal A}_B=\langle\psi|-i\hbar\nabla\psi\rangle
\,.
\eeq
We observe that, in principle,  $\rho_c$ could be made positive-definite by enforcing 
\beq
\boldsymbol{\cal M}+\bn\times\bM=\varrho(\bn-\bn\times\boldsymbol{\cal A}_B)
,
\label{constraint}
\eeq 
so that $\rho_c=\varrho=\int|\tilde\chi|^2\,{ \de^3\mu_R}$. However, enforcing this condition is not an immediate step and this is our main goal in the reminder for this section. For this purpose, we will take the variational principle \eqref{VP-HybKvH2} into a form such that the relation \eqref{constraint} can be directly inserted.

By continuing to proceed analogously to Section \ref{sec:KvH-KvN}, we select a vector field of the form \eqref{equivVF}, so that the variational principle \eqref{VP-HybKvH2} becomes
\beq
\delta\int^{t_2}_{t_1}\!\bigg(\int\!\Big(\boldsymbol{\cal X}\cdot\bM+\bnu\cdot\boldsymbol{\cal M}+  
\varrho \,\langle\psi,i\hbar \partial _t \psi\rangle
+\big(\varrho\bn-\boldsymbol{\cal M}-\bn\times(\bM+\varrho\boldsymbol{\cal A}_B)\big)\cdot\langle\nabla {\sf H}\rangle\Big)\,\de^3 n-h(\varrho,\psi)\bigg)\,\de t=0
,
\label{VP-HybKvH3}
\eeq
where we have defined the following functional for later purpose:
\beq
h(\varrho,\psi)=\int \!\varrho\,\big\langle\psi\big|{\sf H}\psi
+ \nabla {\sf H}\times\bn \cdot (\boldsymbol{\cal A}_B\psi+i\hbar\,\nabla\psi ) \big\rangle \,\de^3 n.
\label{hybHam}
\eeq

At this point, we observe that setting $\boldsymbol{\cal M}=0$ and $\bM=\varrho\partial_\bn S$ eventually returns an equivalent form of the hybrid KvH
equation \eqref{HybKvHeq}. On the other hand, here we want to be able to replace the relation \eqref{constraint} in such a way that the classical density \eqref{classdens2} is made positive-definite by construction. We achieve this in two steps: first, in analogy to the steps preceding Remark \ref{remark}, we adopt the exact solution
$\boldsymbol{\cal M}=\varrho\bn$
arising from the variational problem \eqref{VP-HybKvH3} and, second, we make the replacement
$\bM\to-\varrho\boldsymbol{\cal A}_B$
 in \eqref{VP-HybKvH3}. These steps take the latter into the form
\beq
\delta\int^{t_2}_{t_1}\!\bigg(\int\!\varrho\big(\bnu\cdot\bn+
 \langle\psi,i\hbar (\partial _t +\boldsymbol{\cal X}\cdot\nabla)\psi\rangle\big)\,\de^3 n-h(\varrho,\psi)\bigg)\,\de t=0
\,,
\label{VP-HybKvH4}
\eeq
where the functional \eqref{hybHam} acquires the meaning of a Hamiltonian functional identifying the total energy of the system. Indeed, we notice that if ${\sf H}$ is a purely classical function ${\sf H}(\bn)=H(\bn)\boldsymbol{1}$, then \eqref{hybHam} reduces to the classical total energy $\int \varrho H\de^3 n$, while the case $\nabla{\sf H}=0$ returns the usual expression of the quantum energy $\operatorname{Tr}(\hat\rho_q{\sf H})$, where $\hat\rho_q=\int\varrho\psi\psi^\dagger\de^3 n$ is the density matrix. 
\rem{ 
The appearance of the material derivative $\partial _t +\boldsymbol{\cal X}\cdot\nabla$ in \eqref{VP-HybKvH4} motivates us to express the quantum dynamics of $\psi$ in the phase-space frame moving with velocity $\boldsymbol{\cal X}$. See \cite{} for how this approach was used in the case of classical canonical coordinates. If we let $U(\bn,t)$ be a unitary operator on $\Bbb{C}^2$ parameterized by $\bn$, we can write $\psi(\bn,t)=U(\bn_0,t)\psi_0(\bn_0)|_{\bn_0=\boldsymbol\eta^{-1}(\bn,t)}$, where $\boldsymbol\eta(\bn,t)$ is the phase-space path obeying $\partial_t\boldsymbol\eta(\bn,t)=\boldsymbol{\cal X}(\boldsymbol\eta(\bn,t),t)$. Then, upon introducing the convenient notation 
\[
D=\varrho
,\qquad
\rho=\psi\psi^\dagger
,\qquad
{\bX}_{\sf H}=\nabla {\sf H}\times\bn
,\qquad\text{and}\qquad
\xi(\bn,t)=\partial_t{U}(\bn_0,t)U^{-1}(\bn_0,t)|_{\bn_0=\boldsymbol\eta^{-1}(\bn,t)}
,
\]
 the variational principle \eqref{VP-HybKvH4} becomes
\beq
\delta\int^{t_2}_{t_1}\!\!\bigg(\int\!D\big(\bnu\cdot\bn+
 \langle\rho,i\hbar \xi\rangle\big)\,\de^3 n-h(D,\rho)\!\bigg)\de t=0
,
\quad\ \text{with}\quad\ 
h(D,\rho)=
\int\! D\Big\langle {\sf H}
+\frac{i\hbar}{2} [\nabla\rho,\bX_{{\sf H}}]\Big\rangle\,\de^3 n
\,.
\label{VP-HybKvH5}
\eeq
}  

Before concluding this section, we should emphasize that setting $\boldsymbol{\cal M}=\varrho\bn$ in 
 \eqref{VP-HybKvH3} and  replacing $\bM\to-\varrho\boldsymbol{\cal A}_B$ in the resulting Lagrangian $\int\!\big(\boldsymbol{\cal X}\cdot\bM+\varrho\bnu\cdot\bn+  
\varrho\, \langle\psi,i\hbar \partial _t \psi\rangle
-(\bn\times\bM+\varrho\bn\times\boldsymbol{\cal A}_B)\cdot\langle\nabla {\sf H}\rangle\big)\de^3 n-h(\varrho,\psi)$  has the crucial property of making the latter invariant under classical phase transformations. While this statement appears obvious since phases were completely eliminated in \eqref{VP-HybKvH4}, its meaning is very fundamental: by making our Lagrangian phase invariant, we have made our theory insensitive to classical phases. In other words \cite{Sudarshan}, classical phases are \emph{unobservable}, that is they cannot be measured and indeed cannot contribute to expectation value dynamics. Here, this property has been enforced by resorting to a gauge principle that is directly analogous to the usual phase-invariance of standard quantum mechanics: wavefunctions are defined only up to global phase factors. 
As a result of this gauge-invariance (or independence), the classical density now obeys the transport equation $\partial_t\varrho+\operatorname{div}(\varrho\boldsymbol{\cal X})=0$ so that classical states are identified by a positive density at all times. The next section presents the explicit form of the nonlinear quantum-classical model arising from the  variational principle \eqref{VP-HybKvH4}.

\subsection{Nonlinear model II: quantum-classical equations of motion\label{sec:NlinModII}}
The equations of motion for mixed quantum-classical spin systems can now be obtained by taking the variations in \eqref{VP-HybKvH4}.  In the variational principle \eqref{VP-HybKvH4}, $\delta\psi$ is arbitrary, while $\delta \boldsymbol{\cal X}$ and $\delta \varrho$ are given as in \eqref{XDvar} (upon replacing $D$ by $\varrho$). In addition, $\delta\bnu$ is given in \eqref{xivar} and we recall the auxiliary equation 
\beq
\partial_t\varrho+\operatorname{div}(\varrho\boldsymbol{\cal X})=0
\,.
\label{classevol}
\eeq
Then, upon introducing  the local density matrix $P=\psi\psi^\dagger$, variations $\delta\psi$ yield
\begin{align}\nonumber
i\hbar (\partial _t +\boldsymbol{\cal X}\cdot\nabla)\psi=
&\, 
\frac1{2\varrho}\frac{\delta h}{\delta\psi}
\\
=
&\,
{\sf H}\psi+
{i\hbar}\bigg(\{P,{\sf H}\}+\{{\sf H},P\}+\frac1{2\varrho}\big[P,\{ \varrho,{{\sf H}}\}\big]\bigg)\psi
\,,
\label{quantevol}
\end{align}
where $\delta h/\delta\psi$ denotes the usual functional derivative so that, for example, $\delta F=\int\langle\delta F/\delta\psi,\delta\psi\rangle\,\de^3n$ for any functional $F(\psi)$.
The second equality in \eqref{quantevol} follows from \eqref{hybHam} and by recalling \eqref{LPB}. In particular, 
if we denote $\bX_{\sf H}=\nabla{\sf H}\times\bn$ and $\langle{\sf A}\rangle = \operatorname{Tr}(P{\sf A})$, we notice that \eqref{hybHam} may be rewritten as 
\beq\label{hamfun1}
h(\varrho,P)=
\int \varrho\,\Big\langle {\sf H}
+\frac{i\hbar}2 [\nabla P,\bX_{{\sf H}}]\Big\rangle\,\de^3 n
\,,
\eeq
so that $\delta h/\delta P=\varrho{\sf H}+
{i\hbar}\varrho(\{ P,{\sf H}\}+\{{\sf H}, P\}+[ P,\{\ln \varrho,{{\sf H}}\}]/2)$ and  ${\delta h}/{\delta\psi}=2({\delta h}/{\delta P})\psi$. The Hamiltonian functional \eqref{hamfun1} will be useful for later purpose; see Section \ref{sec:bktstr}.
At this point, the quantum-classical model comprises equations \eqref{classevol} and \eqref{quantevol} while we still need to find the expression of the vector field $\boldsymbol{\cal X}$. 
Variations $\delta\boldsymbol{\cal X}$ in \eqref{VP-HybKvH4} lead to
\begin{align}
\nonumber
\bnu+\nabla\bigg(\frac{\delta h}{\delta \varrho}-\frac1{2\varrho} \bigg\langle \frac{\delta h}{\delta\psi},\psi\bigg\rangle\bigg)
=&\, 
\partial_t\boldsymbol{\cal A}_B+\nabla(\boldsymbol{\cal X}\cdot\boldsymbol{\cal A}_B)-{\boldsymbol{\cal X}}\times\operatorname{curl}\boldsymbol{\cal A}_B
\\
=&\, 
\frac1{\varrho} \Big\langle\frac{\delta h}{\delta\psi},\nabla\psi\Big\rangle-\nabla\left\langle\frac1{2\varrho} \frac{\delta h}{\delta\psi},\psi\right\rangle
\label{anna}
,
\end{align}
where the second equality follows from \eqref{quantevol}. Eventually, we obtain
\[
\bnu=-\nabla\frac{\delta h}{\delta \varrho}+\frac1{\varrho} \bigg\langle\frac{\delta h}{\delta\psi},\nabla\psi\bigg\rangle
\,,
\]
while variations $\delta\bnu$ yield
\beq
{\boldsymbol{\cal X}}=-\bnu\times\bn=\nabla\frac{\delta h}{\delta \varrho}\times\bn-\frac1{\varrho} \bigg\langle\frac{\delta h}{\delta\psi},\nabla\psi\bigg\rangle\times\bn
\,.
\label{elena}
\eeq
Here, ${\delta h}/{\delta\psi}$ is given in \eqref{quantevol} and ${\delta h}/{\delta \varrho}=\left\langle {\sf H}
+{i\hbar} [\nabla P,\bX_{{\sf H}}]/2\right\rangle$. 
Since $\langle{\delta h}/{\delta\psi},\nabla\psi\rangle=\langle{\delta h}/{\delta P},\nabla P\rangle$,  repeated use of the the Leibniz product rule yields
\begin{align}\nonumber
{\boldsymbol{\cal X}} =&\ 
\bX_{{\delta h}/{\delta \varrho}}-\frac1{\varrho} \bigg\langle\frac{\delta h}{\delta P},\bX_P\bigg\rangle
\\
=&\ 
\langle\bX_{{\sf H}}\rangle-\frac{\hbar}{2\varrho}\operatorname{Tr}\!\big( (\varrho\,\boldsymbol{\Xi})\cdot\nabla\bX_{{{\sf H}}}
- \bX_{{{\sf H}}}\cdot\nabla (\varrho\,\boldsymbol{\Xi})
\big)
,\quad\text{where}\quad
\boldsymbol{\Xi}=i
[ P,\nabla P]\times\bn
.
\label{hybvectfield}
\end{align}
Thus, the nonlinear quantum-classical spin model comprises equations \eqref{classevol}, \eqref{quantevol}, and \eqref{hybvectfield}. 
Perhaps not surprisingly, we have formally obtained an equivalent system to \eqref{HybEq1}-\eqref{HybEq3}, provided one replaces $J\widehat{\boldsymbol{\Gamma}}\to\boldsymbol{\Xi}$, $ D\to\varrho$, and the canonical bracket by the Lie-Poisson bracket in \eqref{LPB}. This is a further indication of the level of generality of the proposed variational combination of Koopman wavefunctions and classical phase invariance. Indeed, this approach extends to arbitrary noncanonical systems beyond spin dynamics, although this aspect is beyond the scope of this paper and will be developed in future work.

In the  case of a quantum subsystem comprising a single $1/2-$spin variable, a simple possible Hamiltonian has the form $ {\sf H}( \mathbf{n} )=H_0( \mathbf{n} )+H_I( \mathbf{n} )\widehat{\sigma}_x +\gamma\widehat{\sigma}_z$, where $\gamma$ is a constant parameter. Then, the equations for the classical density and the (local) quantum density matrix,
\[
\partial_t \varrho+\operatorname{div}(\varrho\boldsymbol{\cal X})=0
\,,\qquad\ 
i\hbar(\partial_tP+\boldsymbol{\cal X}\cdot\nabla P)=[{\cal H},P]\,,
\] 
are completed by the following expressions:
\[
{\boldsymbol{\cal X}} 
=
\bX_{H_0}+\bX_{H_I}\langle\widehat\sigma_x\rangle-\frac{\hbar}{2\varrho}\Big( (\varrho\,\widetilde{\!\boldsymbol{\varXi}})\cdot\nabla\bX_{H_I}
-(\bX_{H_I}\cdot\nabla) (\varrho\,\widetilde{\!\boldsymbol{\varXi}})
\Big)
,\quad\text{where}\quad
\,\widetilde{\!\boldsymbol{\varXi}}=i\bn\times\operatorname{Tr}(\nabla P
[ P,\widehat\sigma_x])
\,,
\]
and 
\[
\mathcal{H}=
H_I\widehat{\sigma}_x+\gamma\widehat{\sigma}_z+
{i\hbar}\{[P,\widehat{\sigma}_x],{H_I}\}+\frac{i\hbar}{2\varrho}[P,\widehat{\sigma}_x]\{ \varrho,{{ H_I}}\}.
\]
Notice that, even in the case $\gamma=0$, the last two terms above prevent the  condition $\langle\widehat\sigma_x\rangle=0$ for the local expectation of $\widehat\sigma_x$ from being preserved in time. Indeed, one has $\partial_t\langle\widehat\sigma_x\rangle+{\boldsymbol{\cal X}} \cdot\nabla\langle\widehat\sigma_x\rangle=\gamma\langle\widehat\sigma_y\rangle+\varrho^{-1}\{H_I,\varrho(\langle\widehat\sigma_x\rangle^2-1)\}$. This is in contrast with the predictions resulting from the simpler Ehrenfest model, which is obtained by neglecting all the $\hbar-$terms in ${\boldsymbol{\cal X}} $ and $\mathcal{H}$. In the case when $\gamma=0$, we observe that the $\varrho-$equation produced by the Ehrenfest model decouples completely from the quantum evolution.

\section{Discussion\label{sec:discussion}}

\subsection{Hybrid density and energy conservation}

This section considers the nature of the total energy expressed by the Hamiltonian functional in \eqref{hybHam}. While the first term is well known from Ehrenfest dynamics,  the $\hbar-$term in \eqref{hybHam} is only triggered by the presence of quantum-classical coupling terms in ${\sf H}$ and its nature is currently a matter of speculation. In canonical coordinates \cite{GBTr22}, that is in the case of orbital degrees of freedom in the classical sector,  this term was related to the fluctuation force $\widetilde{F}=-\partial_q V(q,x)+\langle \partial_q V(q,x)\rangle$
 that arises from the interaction potential $V(q,x)$. Indeed, in this case this $\hbar-$term becomes $\int\varrho\,\langle\psi,-i\hbar\widetilde{F}\cdot\partial_p\psi\rangle\,\de q\de p$. Analogously, in the present case we may define the fluctuation torque $\widetilde{\bf T}=(\nabla{\sf H}-\langle\nabla{\sf H}\rangle)\times\bn$, so that \eqref{hybHam} becomes
\[
h(\varrho,\psi)=\int\!\varrho\big(\langle{\sf H}\rangle-\langle\psi,i\hbar\widetilde{\bf T}\cdot\nabla\psi\rangle\big)\de^3 n
.
\] 
However, the deeper meaning of the second term in the integral remains an open question. 

One possible interpretation, which follows the ideas in \cite{carroll}, is to accept that quantum-classical coupling does not conserve energy in the usual sense, as given in the first term of the integral above, thereby involving some kind of energy transfer that is triggered by quantum-classical correlations. Motivated by speculations on the measurement problem, this interpretation would explain the additional work term as a model for this extra quantum-classical energy transfer.

If, on the other hand, we insist that the physical energy must be conserved, then we can integrate by parts in the integral above and follow the procedure outlined in Section \ref{sec:hybmod1}. In particular, we verify that the Hamiltonian functional  $h(\varrho,\psi)$ may be rewritten as $\operatorname{Tr}\int {\sf H}\mathcal{D}\,\de^3 n$ where
\[
{\cal D}
= 
\varrho{P}+\frac{\hbar}2\operatorname{div}\!\left(\varrho\boldsymbol{\cal J}\times\bn\right)=\varrho{P}+\frac{\hbar}2\bn\cdot\operatorname{curl}(\varrho\boldsymbol{\cal J})
,\qquad\text{with}\qquad
\boldsymbol{\cal J}=i[P,\nabla P].
\]
Here, the quantity $\boldsymbol{\cal J}$ is a non-Abelian gauge potential already appeared in Mead's work on geometric phases \cite{Me92}. 
Then, we may interpret $\operatorname{Tr}\int {\sf H}\mathcal{D}\,\de^3 n$ as the expectation value of the quantum-classical Hamiltonian ${\sf H}$, which is computed in the usual way by using the above distribution-valued Hermitian operator. However, if we follow this interpretation, we should notice that, unlike ${\cal P}=\varrho P$, the operator ${\cal D}$ is generally unsigned and thus its meaning requires further considerations. A possible way of discerning which of the two interpretations is more suitable consists in comparing the expected value dynamics that is computed by using $\cal P$ and $\cal D$. If the results arising from using $\cal D$ turn out unrealistic, then the first interpretation could be more appropriate and one is led to revise the concept of energy conservation. This investigation needs adequate computational efforts, which are currently under development.

Here, we point out that classical and quantum expectation values are unaffected by the interpretation of $\cal D$. Indeed, we notice that the classical Liouville density and the quantum density matrix are given respectively  as
\[
\rho_c=\operatorname{Tr}{\cal D}=\varrho
,\ \qquad\text{ and }\qquad\ 
\hat{\rho}_q=\int\!{\cal D}\,\de^3 n=\int\!{\cal P}\,\de^3 n,
\]
where we have used $\operatorname{Tr}{\cal P}=\varrho$. 
Given a classical and a quantum observable, $A(\bn)$ and ${\sf A}$, respectively, their expectation values are  
\[
\int A(\bn)\varrho(\bn)\,\de^3 n,\ \qquad\text{ and }\qquad\ \operatorname{Tr}({\sf A}\hat{\rho}_q),
\]
which involve only the operator ${\cal P}=\varrho P$ rather than the entire expression of $\cal D$. The equations of motion of these expectation value require an extension of the Ehrenfest theorem to the quantum-classical setting. As presented in \cite{GBTr21}, this extension is a consequence of the covariance properties of the $\cal D$ operator. Indeed, upon writing ${\cal D}[{\cal P}]={\cal P}-(i\hbar/2)\operatorname{div}\!\left(\bn\times[{\cal P},\nabla{\cal P}]/\operatorname{Tr}{\cal P}\right)$ a direct application of the methods in \cite{GBTr21} yields the following covariance properties
\[
\widehat{\cal D}[{\cal P}( {\boldsymbol\upeta}(\bn))]= \widehat{\cal D}[{\cal P}] ( {\boldsymbol\upeta}(\bz))
\qquad\text{and}\qquad
\widehat{\cal D}[{\cal U}{\cal P}{\cal U}^\dagger]={\cal U}\widehat{\cal D}[{\cal P}]{\cal U}^\dagger.
\]
Here, ${\cal U}$ is a quantum unitary operator while $\boldsymbol\upeta$ is a classical canonical transformation (more exactly, a \emph{Poisson diffeomorphism}, since we are dealing with noncanonical coordinates). We emphasize that the covariance properties of $\cal D$ have long been sought in classical-quantum theories and they correspond to property 3 in the list of quantum-classical consistency  properties presented in Section \ref{sec:intro}.

As showed in \cite{GBTr21}, these covariance properties lead to casting the equations for the classical density  and the quantum density matrix  as
\beq\label{densevol2}
\frac{\partial\rho_c}{\partial t}=\operatorname{\sf Tr}\{{\sf H},{\cal D}\}
\,,\qquad\qquad
i\hbar\frac{\de\hat\rho_q}{\de t}=\int[{\sf H},{\cal D}]\,\de^3 n
\,,
\eeq
which are formally identical to the analogous relations \eqref{rho_c} and \eqref{rho_q} in the original KvH treatment. For example, the first in \eqref{densevol2} can be  verified upon rewriting the vector field \eqref{hybvectfield} as
\[
{\boldsymbol{\cal X}} =\frac1\varrho\,\langle{\cal D}, \bX_ {\sf H}\rangle+\frac\hbar\varrho\operatorname{div}\!\big(\varrho\operatorname{Tr}\!\big( \bX_ {\sf H}\wedge\,\boldsymbol{\Xi}\big)\big)
,\qquad \text{where}\qquad
\big( \bX_ {\sf H}\wedge\,\boldsymbol{\Xi}\big)^{jk}=\frac12\big( X_ {\sf H\,}^j{\Xi}^k-{\Xi}^j X_ {\sf H}^k\big)
\]
is a skew-symmetric contravariant two-tensor, that is a \emph{bivector}.  
Then, the hybrid Ehrenfest dynamics of classical and quantum expectation values arises as a direct consequence of  \eqref{densevol2}.

\subsection{Bracket structure\label{sec:bktstr}}

In this section, we present the quantum-classical bracket structure underlying the dynamical model comprised by equations \eqref{classevol}, \eqref{quantevol}, and \eqref{hybvectfield}. The search for hybrid brackets in quantum-classical dynamics was first motivated by the observation that the right-hand side of the Aleksandrov-Gerasimenko (AG) equation \cite{Aleksandrov,Gerasimenko} $\partial_t{\cal P}=-i\hbar^{-1}[{\sf H},{\cal P}]+(\{{\sf H},{\cal P}\}+\{{\cal P},{\sf H}\})/2$ defines a bracket structure that does not satisfy the Jacobi identity \cite{Se05}. Despite its popularity  in quantum chemistry \cite{Kapral,SuOuLa13}, the AG equation fails to retain positivity of the quantum density matrix $\hat\rho_q=\int {\cal P}\,\de q \de p$ at all times, thereby invalidating Heisenberg's uncertainty principle. Nevertheless, several studies are still based on the AG equation and this has led to some speculations on whether quantum-classical mechanics may at all have a Hamiltonian formulation \cite{Salcedo}. For theories involving a classical phase-space description, Poisson bracket structures were recently provided by the authors in \cite{GBTr22,GBTr21}, where we showed that our current model actually possesses the same Poisson bracket as the Ehrenfest model  \eqref{Ehrenfest}. While this Poisson structure was only provided for classical orbital degrees of freedom (canonical coordinates), here we present its extension to quantum-classical spin systems.

As we observed above, the passage from canonical to noncanonical coordinates merely requires replacing the canonical Poisson bracket $\{F,G\}_{\tiny\rm can\!}=\partial_qF\partial_pG-\partial_pF\partial_qG$ by the Lie-Poisson bracket $\{F,G\}=\bn\cdot\nabla F\times\nabla G$. This rule applies also to the bracket structure for the quantum-classical model in equations \eqref{classevol}, \eqref{quantevol}, and \eqref{hybvectfield}. In order to see this, it is convenient to express the entire quantum-classical model in terms of the density-valued operator ${\cal P}=\varrho P$ so that $\langle A \rangle=\operatorname{Tr}({\cal P}A)/\operatorname{Tr}{\cal P}$. First, upon recalling the notation $\bX_{A}=\nabla{A}\times\bn$, we  rewrite the first line of \eqref{hybvectfield} as 
${\boldsymbol{\cal X}} =
 \langle\bX_{\delta h/\delta\cal P}\rangle$, where ${\delta h}/{\delta {\cal P}}=({\delta h}/{\delta P}-\langle{\delta h}/{\delta P}\rangle\boldsymbol{1})/\varrho+({\delta h}/{\delta \varrho})\boldsymbol{1}$
and the Hamiltonian \eqref{hamfun1} now reads
 \beq
h({\cal P})
=\int \big\langle{\sf H}\operatorname{Tr}{\cal P}
+  {i\hbar}\{{\cal P}, {\sf H}\}\big\rangle  \,\de^3 n.
\label{hybHam2}
\eeq
In addition, we notice that, by the relation ${\delta h}/{\delta\psi}=2({\delta h}/{\delta P})\psi$, the first line in \eqref{quantevol} may be written as $i\hbar \varrho(\partial _t +\boldsymbol{\cal X}\cdot\nabla)P=
[{\delta h}/{\delta P},P]$. Thus, since $[\delta h/\delta P,P]=[\delta h/\delta {\cal P},{\cal P}]$, the equations \eqref{classevol} and \eqref{quantevol} are equivalently rewritten in Hamiltonian form as 
\[
i\hbar\frac{\partial{\cal P}}{\partial t} + i\hbar\operatorname{div}\!\Big({\cal P}\Big\langle\bX_\frac{\delta h}{\delta\cal P}\!\Big\rangle\Big)=
\bigg[\frac{\delta h}{\delta {\cal P}},{\cal P}\bigg], 
\]
where  the functional derivative of \eqref{hybHam2} is given by
\[
\frac{\delta h}{\delta {\cal P}}={\sf H}+
\frac{\hbar}{2\operatorname{Tr}{\cal P}}\Big(2i(\{{\cal P},{\sf H}\}+\{{\sf H},{\cal P}\})+i[{\cal P},\{{\sf H},\ln \operatorname{Tr}{\cal P}\}]-\langle i\{{\cal P},{\sf H}\}+i\{{\sf H},{\cal P}\}\rangle\boldsymbol{1}\Big).
\]
If we now introduce the convenient notation $A\!\!:\!\!B=\operatorname{Tr}(AB)$, the usual relation $\dot{f}=\{\!\!\{f,h\}\!\!\}$  holding for any functional of the dynamical variables ($\cal P$ in this case) yields
\begin{align}\nonumber
\{\!\!\{f,h\}\!\!\}(\mathcal{P})
&\
= 
\int \!\frac1{\operatorname{Tr}\mathcal{P}  }\bigg(\mathcal{P}  \!:\! \left\{\frac{\delta f}{\delta \mathcal{P}},\frac{\delta h}{\delta \mathcal{P}}\right\}\!:\! \mathcal{P}   \bigg)  \,\de^3 n
-\int  \!\left\langle \mathcal{P}  ,\frac{i}\hbar\!\left[\frac{\delta f}{\delta \mathcal{P}},\frac{\delta h}{\delta \mathcal{P}}\right] \right\rangle\de^3 n 
\\
&\ 
=
\int  \!\bigg(
 \bn\cdot\bigg\langle\nabla\frac{\delta f}{\delta \mathcal{P}}\bigg\rangle\times\bigg\langle\nabla\frac{\delta h}{\delta \mathcal{P}}\bigg\rangle
 -
 \left\langle \frac{i}\hbar\!\left[\frac{\delta f}{\delta \mathcal{P}},\frac{\delta h}{\delta \mathcal{P}}\right] \right\rangle\!
   \bigg) {\operatorname{Tr}\mathcal{P}  }\, \de^3 n 
   \,.
 \label{bracket_candidate_rho}
\end{align} 
This bracket can be given an equivalent expression in terms of $\varrho$ and $\mathcal{P}$. The explicit proof that the bracket \eqref{bracket_candidate_rho} is Poisson involves several technical steps in Poisson geometry and is left for another venue. In the case of classical canonical coordinates, the corresponding proof was presented explicitly in  \cite{GBTr21}, to which we refer for a detailed discussion. Here, we will simply say that the above bracket is Poisson because it was obtained from a variational principle of the particular type \eqref{VP-HybKvH4}, which can be shown to identify a Hamiltonian system for  arbitrary functionals $h(\varrho,\psi)$. Then, together with the Hamiltonian functional \eqref{hybHam2}, the Poisson bracket \eqref{bracket_candidate_rho} comprises the Hamiltonian structure of the quantum-classical spin model given by the equation
\beq\label{maria}
i\hbar\partial_t{\cal P}+i\hbar\operatorname{div}(\boldsymbol{\cal X}{\cal P})=[{\cal H},{\cal P}],
\eeq
and the definitions
\begin{align}
\boldsymbol{\cal X}&=
\langle\bX_{{\sf H}}\rangle-\frac{\hbar}{2\varrho}\operatorname{Tr}\!\big( (\varrho\,\boldsymbol{\Xi})\cdot\nabla\bX_{{{\sf H}}}
- \bX_{{{\sf H}}}\cdot\nabla (\varrho\,\boldsymbol{\Xi})
\big)\label{mathcalX}
\\
{\cal H}&
=
{\sf H}+\frac{i\hbar}{\varrho}\Big(\{\mathcal{P},{{\sf H}}\}+\{{\sf H},\mathcal{P}\} +\frac12[\{\ln \varrho,{\sf H}\},\mathcal{P}] 
\Big),\label{mathcalH}
\end{align}
where $\boldsymbol{\Xi}=i
[ {\cal P},\nabla {\cal P}]\times\bn/\varrho^2$ and $\varrho=\operatorname{Tr}{\cal P}$. It is easy to verify that  the relation ${\cal P}=\varrho\psi\psi^\dagger$ indeed yields the  equations \eqref{classevol}, \eqref{quantevol}, and \eqref{hybvectfield}. Notice that a factor $\varrho^{-1}$ has been taken inside the final Poisson bracket $\{\, , \}$ for convenience of notation, thereby leading to the appearance of the logarithm $\ln\varrho$. More importantly, we remark that a direct expansion of $\operatorname{div}\!\boldsymbol{\cal X}$ leads to observing that no ${\cal P}-$gradients of order higher than two are involved in the equation \eqref{maria}.

\subsection{Invariants of motion\label{sec:invariants}}
Given the quantum-classical bracket \eqref{bracket_candidate_rho}, one may ask wether this is accompanied by Casimir invariant functionals $C({\cal P})$ such that $\dot{C}=\{\!\!\{C,f\}\!\!\}=0$ for any functional $f({\cal P})$. This question was addressed in \cite{GBTr21} for the case of classical orbital degrees of freedom and here we present the noncanonical spin counterpart. Given the structure of the evolution equation \eqref{maria} for $\cal P$ it is not difficult to see that the functional 
\[
C=\operatorname{Tr}\int\!\varrho\Phi({\cal P}/\varrho)\,\de^3 n
,\qquad\text { with }\qquad
\varrho=\operatorname{Tr}{\cal P}
\]
identifies a dynamical invariant for any  analytic matrix function $\Phi$. For example, one can construct the entropy-like invariant $S=-\operatorname{Tr}\int{\cal P}\log({\cal P}/\varrho)\,\de^3 n$, which reduces to the von-Neumann entropy $-\operatorname{Tr}(\hat\rho_q\log\hat\rho_q)$ when ${\cal P}=\varrho\hat\rho_q$. See \cite{GBTr22} for further remarks on entropy functionals in quantum-classical mechanics.

Other Casimir invariants may also be constructed for ${\cal P}=\varrho\psi\psi^\dagger$ upon noticing that the dynamics may be restricted to occur on functions defined on the unit sphere by setting $\|\bn\|=1$, as we will now discuss. First, from \eqref{elena} we observe that the vector field $\boldsymbol{\cal X}$ is tangent to the $S^2$. Furthermore, we can show that the expression of the vector field $\boldsymbol{\cal X}$ naturally restricts to the unit sphere $S^2$.
To see this, we first recall that $\delta h/\delta {\cal P}$ is a function of $(\varrho, {\cal P},{\sf H})$ and $(\nabla\varrho,\nabla {\cal P},\nabla{\sf H})$. Now due to its expression in \eqref{mathcalH}, we  have that $\delta h/\delta {\cal P}$ is uniquely expressed as a function of $(\varrho,{\cal P},{\sf H})|_{n=1}$ and $(\mathscr{D}(\varrho|_{n=1}),\mathscr{D} ({\cal P}|_{n=1}),\mathscr{D}({\sf H}|_{n=1}))$, where $\mathscr{D}(\,\cdot\,|_{n=1}):=(\boldsymbol{1}-\bu\bu^T)\nabla(\cdot)|_{n=1}$ is the gradient on $S^2$ and  we have denoted $\bn=n\bu$, with $\|\bu\|=1$. This follows from the fact that \eqref{mathcalH} involves the Lie-Poisson bracket in \eqref{LPB}, which is known to restrict to the unit sphere $S^2$, this restriction being obtained by evaluating at $n=1$ and replacing $\nabla(\cdot)\to(\boldsymbol{1}-\bu\bu^T)\nabla(\cdot)|_{n=1}$. From this, and from the fact that the expression of the Lie-Poisson vector field $\bX_H=\nabla H\times\bn$ also restricts to the sphere by the same process, we obtain that the same holds for $\boldsymbol{\cal X}=\langle\bX_{\delta h/\delta{\cal P}}\rangle$ in \eqref{mathcalX}.
As a consequence, we have
\[
\operatorname{div}(\boldsymbol{\cal X}{\cal P})|_{n=1}=\operatorname{Div}(\boldsymbol{\cal X}{\cal P}|_{n=1}),
\]
where we notice that $\operatorname{Div}(\boldsymbol{\cal X}{\cal P}|_{n=1})=\operatorname{div}(\boldsymbol{\cal X}{\cal P})|_{n=1}-\bu\cdot\nabla(\boldsymbol{\cal X}{\cal P})|_{n=1}\cdot\bu$ is the expression of the divergence on the sphere. Therefore, the restriction ${\cal P}|_{n=1}=\varrho\psi\psi^\dagger|_{n=1}$ satisfies exactly the same equation \eqref{maria}  upon replacing $\bn\to\bu$ and $\operatorname{div}(\cdot)\to\operatorname{Div}(\,\cdot\,|_{n=1})$. For later purpose, here we also notice that $\bu\times{\boldsymbol{\cal X}}|_{n=1}=-(\boldsymbol{1}-\bu\bu^T)\left({\varrho}^{-1} \langle{\delta h}/{\delta\psi},\nabla\psi\rangle-\nabla({\delta h}/{\delta \varrho})\right)|_{n=1}$.

As anticipated, the fact that the entire quantum-classical dynamics restricts naturally to the unit sphere allows us to identify new Casimir invariants. In order to proceed, as we are now on a manifold, it is convenient to use differential forms and the Lie derivative $\pounds$. In particular, upon using Cartan's magic formula, we have $\pounds_{\boldsymbol{\cal X}}\omega=\de(\mathbf{i}_{\boldsymbol{\cal X}}\omega)$ for any two-form $\omega$ on $S^2$, where we have replaced $\boldsymbol{\cal X}|_{n=1}\to\boldsymbol{\cal X}$ for convenience of notation. In the particular case of the area form on $S^2$, we have $\omega_{jk}(\bu)=-\hat{u}_{jk}$ so that $\pounds_{\boldsymbol{\cal X}}\omega=\de(\bu\times{\boldsymbol{\cal X}})$. On the other hand, we may also restrict the second equality in \eqref{anna} thereby obtaining
\[
(\partial_t+\pounds_{\boldsymbol{\cal X}})\boldsymbol{\cal A}_B
=
\frac1{\varrho} \Big\langle\frac{\delta h}{\delta\psi},\de\psi\Big\rangle-\de\left\langle\frac1{2\varrho} \frac{\delta h}{\delta\psi},\psi\right\rangle=-\bu\times{\boldsymbol{\cal X}}-\de\!\left(\frac{\delta h}{\delta \varrho}+\left\langle\frac1{2\varrho} \frac{\delta h}{\delta\psi},\psi\right\rangle\right)
\]
where $\boldsymbol{\cal A}_B=\langle\psi,-i\hbar\de\psi\rangle$ and we have replaced the notation $(\varrho|_{n=1},\psi|_{n=1})\to(\varrho,\psi)$. Taking the differential on both sides and defining ${\cal B}=\de \boldsymbol{\cal A}_B=\hbar\operatorname{Im}\{\psi^\dagger,\psi\}\omega$  yields 
\[
(\partial_t+\pounds_{\boldsymbol{\cal X}})(\omega+{\cal B})=0
\,.
\]
At this point, for any scalar function $\Phi({\sf x})$,  we obtain the following Casimir invariant
\[
C=\iint_{S^2}\varrho\Phi({\Omega}/\varrho)\omega
\,,\quad\ \ \text{where}\quad\ \ 
\Omega:=1+\hbar\operatorname{Im}\{\psi^\dagger,\psi\}.
\]
For example, we have the following relation
\[
\frac{\de }{\de t}\iint_{S^2}\varrho\log({\Omega}/\varrho)\omega=0
\,.
\]
Notice that in the classical case the Berry connection is exact \cite{GBTr22} so that ${\cal B}=0$ and the Casimir $\iint_{S^2}\varrho\log({\Omega}/\varrho)$ recovers the usual Gibbs entropy functional.

\subsection{Spin systems and orbital degrees of freedom}

Given the formal similarities between the cases of orbital and spin degrees of freedom in the classical sector, we are motivated to propose  extensions of the present model to more general situations. For example, realistic spin systems involve several quantum and classical spins, so that the hybrid Hamiltonian $\mathsf{H}$, resp., the hybrid operator ${\cal P}$, becomes a function, resp., a density, on $j$ classical spin variables $(\bn_1,\dots\bn_j)$ taking values in the space of Hermitian operators on the  Hilbert space $(\Bbb{C}^2)^{\otimes k}$ for $k$ quantum spins. Then, the classical density $\varrho=\operatorname{Tr}{\cal P}$ involves a generalized trace on the full tensor-product space. The quantum-classical model presented here extends naturally to this situation. In particular, the hybrid density ${\cal P}(\bn_1,\dots\bn_j)$ obeys the equation
\beq\label{hybeqnfinal}
i\hbar\partial _t{\cal P} + i\hbar\operatorname{div}({\cal P}\boldsymbol{\cal X})=
[{\cal H},{\cal P}]
\,,
\eeq
with
\[
\boldsymbol{\cal X}=
\langle\bX_{{\sf H}}\rangle-\frac{\hbar}{2\varrho}\operatorname{Tr}\!\big( (\varrho\,\boldsymbol{\Xi})\cdot\nabla\bX_{{{\sf H}}}
-(\bX_{{{\sf H}}}\cdot\nabla) (\varrho\,\boldsymbol{\Xi})
\big)
,\qquad\text{with}\qquad
\boldsymbol{\Xi}=i\varrho^{-1}
[ {\cal P},\bX_{\cal P}]\,,
\]
and
\[
{\cal H}=
{\sf H}+
\frac{i\hbar}{\varrho}\Big(\{{\cal P},{\sf H}\}+\{{\sf H},{\cal P}\}+\frac12[{\cal P},\{{\sf H},\ln \varrho\}]\Big)
\,.
\]
Here, $\nabla=(\partial_{\bn_1},\dots,\partial_{\bn_j})$, while the Poisson bracket and the Hamiltonian vector field are given respectively as follows:
\[
\{F,G\}=\sum_{a=1}^j\bn_a\cdot\partial_{\bn_a} F\times\partial_{\bn_a} G
,\qquad\text{ and }\qquad
\bX_A=(\partial_{\bn_1} A\times \bn_1,\dots,\partial_{\bn_j} A\times \bn_j).
\]
As we can see, this case of multiple classical and quantum spins does not exhibit essential differences with respect to the simpler case  presented earlier.

Another case of interest in quantum-classical mechanics involves  classical systems comprising {\it both} orbital and spin degrees of freedom. Our treatment extends directly also to this more general case. Indeed, the equation \eqref{hybeqnfinal} for the hybrid density ${\cal P}(\bq,\bp,\bn)$ now involves the definitions 
\[
\boldsymbol{\cal X}=
\langle\bX_{{\sf H}}\rangle-\frac{\hbar}{2\varrho}\operatorname{Tr}\!\big( (\varrho\,\boldsymbol{\Xi})\cdot\nabla\bX_{{{\sf H}}}
-(\bX_{{{\sf H}}}\cdot\nabla) (\varrho\,\boldsymbol{\Xi})
\big)
,\qquad\text{with}\qquad
\boldsymbol{\Xi}=i\varrho^{-1}
[ {\cal P},\bX_{\cal P}]\,,
\]
and
\[
{\cal H}=
{\sf H}+
\frac{i\hbar}{\varrho}\Big(\{{\cal P},{\sf H}\}+\{{\sf H},{\cal P}\}+ \frac12[{\cal P},\{{\sf H},\ln \varrho\}]\Big)
\,.
\]
Here, we notice that $\nabla=(\partial_\bq,\partial_\bp,\partial_\bn)$. Also, we have recalled $\varrho=\operatorname{Tr}\cal P$ and we have used the following notation
\[
\{F,G\}=\partial_\bq F\cdot\partial_\bp G-\partial_\bp F\cdot\partial_\bq G+\bn\cdot\partial_\bn F\times\partial_\bn G
,\qquad\text{ and }\qquad
\bX_A=(\partial_\bp A,-\partial_\bq A,\partial_\bn A\times \bn),
\]
for the Poisson bracket and the Hamiltonian vector field, respectively.

At this point, one can add different layers of complication depending on the case under consideration. For example, one may also extend to consider a more general quantum Hilbert space other than $ \mathbb{C} ^2$. However, we should emphasize that all the properties discussed previously for elementary quantum-classical spin systems  transfer naturally to these more general settings and in many cases one simply needs to  use the appropriate expression of the Poisson bracket and the Hamiltonian vector field. Given the level of difficulty of these extended versions of the model, we will not pursue this direction  in this paper. For the fully classical Liouville dynamics on the extended phase space with coordinates $(\bq,\bp,\bn)$, we refer the reader to \cite{GiHoKu83}. For applications to complex fluids, see \cite{GBTr10,Tr12}.

\section{Conclusions and future directions}
The equations \eqref{classevol}, \eqref{quantevol}, and \eqref{hybvectfield} comprise a new model for quantum-classical spin dynamics that overcomes a series of consistency issues previously appeared in the literature \cite{AgCi07,boucher}. Going beyond the well-known Ehrenfest model \eqref{Ehrenfest}, this model appears to be the first to satisfy all the following properties: 1) the classical system is identified by a phase-space probability density at all times; 2) the quantum system is identified by a positive-semidefinite density operator $\hat\rho$ at all times; 3) the model is covariant under both quantum unitary transformations and classical canonical transformations; 4) in the absence of an interaction potential, the model reduces to uncoupled quantum and classical dynamics; 5) in the presence of an interaction potential, the {\it quantum purity} $\|\hat\rho\|^2$ is not a constant of motion. For example, property 2) ensures that the Heisenberg principle is satisfied at all times, while property 5) has a crucial importance in quantum chemistry applications \cite{AkLoPr14,CrBa18}. Most importantly, the $\hbar-$terms appearing in \eqref{quantevol} and \eqref{hybvectfield} have the role of retaining quantum-classical correlations via the appearance of several gradient terms involving both the local operator $P=\psi\psi^\dagger$ and the classical density $\varrho$. These quantum-classical correlations take the present model beyond the Ehrenfest system, whose main limitation  is indeed the failure to retain correlation effects.

The nonlinear model \eqref{classevol}-\eqref{hybvectfield} was obtained by combining the variational setting of Koopman wavefunctions in noncanonical coordinates with a gauge principle that makes classical phases unobservable. This construction is based on a thorough investigation of noncanonical KvH and KvN classical dynamics, which required a careful study starting early on with the variational setting of the  Liouville equation for classical spin systems. Along the way, we were able to relate the variational setting of (prequantum) KvH theory with that of the simpler KvN construction. This step was of crucial importance to obtain the quantum-classical model at a later stage.

 Despite its formidable appearance, the  model \eqref{classevol}-\eqref{hybvectfield} does not involve gradients of order higher than two. In addition, it possesses a number of relevant properties such as an explicit Hamiltonian structure and  infinite families of dynamical invariants. In addition, the expression \eqref{hybHam} of the total energy leads to unexpected energy-balance considerations that seem to connect to recent speculations in measurement theory \cite{carroll}. Alternatively, the usual energy balance is recovered by resorting to a hybrid von Neumann operator, which however appears to be generally unsigned. While this hybrid operator may be used for the calculation of  quantum-classical  expectation values, the calculation of purely quantum and purely classical expectation values involves only the positive operator density ${\cal P}=\varrho\psi\psi^\dagger$. Finally, we showed how the proposed model adapts immediately to quantum-classical systems with many spins, with and without the presence of orbital degrees of freedom.

While the proposed model may require appropriate computational tools, due to the intricate structure of its equations, we believe that  its underlying Hamiltonian and variational structures may stand as a platform for designing new convenient algorithms in mixed quantum-classical dynamics. Indeed, in the context of  nonadiabatic molecular dynamics, ``the parallel development of rigorous but computationally expensive methods and more approximate but computationally efficient methods'' is regarded as absolutely ``critical'' \cite{HaSc22}. For example, current investigations using classical canonical coordinates are devoted to construct variational trajectory methods by resorting to suitable regularizations of the Lagrangian  underlying the model equations. Benefiting from the recent results in \cite{HoRaTr21}, this study is currently underway. 

In addition, we hope that a closer look at the proposed model may lead to a better understanding of the peculiarities of quantum-classical coupling. While many of these aspects are usually studied in the context of measurement theory, the latter usually requires invoking the role of an {\it environment} acting as a heat bath thereby making the entire dynamics irreversible. However, we believe that reversible quantum-classical dynamics may already disclose interesting features and the emergence of questions about the energy balance could indeed be a starting point in this direction.

In the future, we would like to understand how the proposed model may be used to describe radical pairs in organic chemistry \cite{Manolopoulos1}. In this context, electron spin-spin coupling requires a quantum treatment while nuclear spin dynamics may be accommodated by semiclassical descriptions \cite{Manolopoulos2}.

Also, motivated by recent studies in  magnonics \cite{YuCaKaReYa22}, we would like to understand how classical spin waves may be coupled to quantum spin systems. In more generality, classical magnetization dynamics may require Gilbert's dissipative  terms and it would be interesting to design models coupling classical Landau-Lifshitz-Gilbert dynamics with quantum spin systems. This seems to be the general concept  underlying the design of recent quantum control strategies in spintronics \cite{RuKaUp22}. In this case, the idea is  to exploit a new range of control parameters that is offered by coupling quantum spins to classical ferromagnets. All these directions necessitate  mathematically sound dynamical models and we hope that the theory proposed here can help in this endeavor.

\medskip

\paragraph{Acknowledgments.} We are grateful to Irene Burghardt, Francesco Di Maiolo, Darryl Holm, Ilon Joseph, Giovanni Manfredi, and Phil Morrison, for their keen remarks on these results.  This work was made possible through the support of Grant 62210 from the John Templeton Foundation. The opinions expressed in this publication are those of the authors and do not necessarily reflect the views of the John Templeton Foundation. CT also acknowledges partial support by the Institute of Mathematics and its Applications and by the Royal Society Grant IES\textbackslash R3\textbackslash203005.

\end{document}